\documentclass[11pt]{article}
\usepackage[left=1in,right=1in,top=1in,bottom=1in]{geometry} 
\usepackage{graphicx}
\usepackage{amssymb}
\usepackage{amsthm}
\usepackage{float}
\usepackage[round,sort]{natbib}
\usepackage{subcaption}
\usepackage{amsmath,comment,color}
\graphicspath{{Images/}}
\usepackage{bm} 
\usepackage[noabbrev]{cleveref} 
\usepackage{authblk}

\newcommand{\bu}{{\bf u}}

\newcommand{\bB}{{\bf B}}

\newcommand{\rhat}{\hat{\bf r}}

\newcommand{\bn}{{\boldsymbol \nabla}}

\newcommand{\zhat}{\hat{\bf z}}

\newcommand{\curl}{\bn\times}

\newcommand{\overbar}[1]{\mkern 1.5mu\overline{\mkern-1.5mu#1\mkern-1.5mu}\mkern 1.5mu}

\newcommand{\pd}[2]{\dfrac{\partial #1}{\partial #2}} 
\renewcommand{\vec}[1]{\bm{#1}} 
\newcommand{\grad}{\vec{\nabla}}

\newtheorem*{theorem}{Theorem}

\title{Constraints on the magnetic field within a stratified outer core}
\author[1]{Colin M. Hardy}
\author[2]{Philip W. Livermore}
\author[3]{Jitse Niesen}
\affil[1]{EPSRC Centre for Doctoral Training in Fluid Dynamics, University of Leeds, Leeds, LS2 9JT, UK}
\affil[2]{School of Earth and Environment, University of Leeds, Leeds, LS2 9JT, UK}
\affil[3]{School of Mathematics, University of Leeds, Leeds, LS2 9JT, UK}

\begin{document}
\maketitle

\abstract

Mounting evidence from both seismology and experiments on core composition suggests the existence of a layer of stably stratified fluid at the top of Earth's outer core. 
In this work we examine the structure of the geomagnetic field within such a layer, building on the important but little known work of \cite{malkus1979dynamo}.
We assume (i) an idealised magnetostrophic spherical model of the geodynamo neglecting inertia, viscosity and the solid inner core, and (ii) a strongly stratified layer of constant depth immediately below the outer boundary within which there is no spherically radial flow. Due to the restricted dynamics, Malkus showed that the geomagnetic field must obey certain a condition which is a refined and more restrictive version of the well known condition of \citet{Taylor_63} which holds on an infinite set of azimuthal rings within the stratified layer. By adopting a spectral representation with truncation $N$ in each direction, we show that this infinite class collapses to a discrete set of $O(N^2)$ Malkus constraints. Although fewer than the $N^3$ degrees of freedom of the magnetic field, their nonlinear nature 
makes 
finding a magnetic field that obeys such constraints, here termed a {\it Malkus state}, a
challenging task.  
Nevertheless, such Malkus states when constrained further by geomagnetic observations have the potential to probe the interior of the core.

By focusing on a particular class of magnetic fields for which the Malkus constraints are linear, we describe a constructive method that turns any purely-poloidal field into an exact Malkus state by adding a suitable toroidal field. 
We consider poloidal fields following a prescribed smooth profile within the core that match a degree-13 observation-derived model of the magnetic field in epoch 2015 or a degree-10 model of the 10000-yr time averaged magnetic field.
Despite the restrictions of the Malkus constraints, a significant number of degrees of freedom remain for the unknown toroidal field and we seek extremal examples. The Malkus state with the least toroidal energy has in both cases a strong azimuthal toroidal field, about double the magnitude of that observed from the poloidal field at the core-mantle boundary. 
For the 2015 field for a layer of depth 300 km, we estimate a root mean squared azimuthal toroidal field of $3$~mT with a pointwise maximum of 8 mT occurring at a depth of about 70 km.  

\section{Introduction}
The question of whether or not Earth's liquid outer core contains a stratified layer just below its outer boundary has long been debated \citep{whaler1980does,Braginsky_67,braginsky1987waves,hardy2019stably,gubbins2007geomagnetic}. A stratified layer may result from the pooling of buoyant elements released from the freezing of the solid inner core \citep{braginsky2006formation,bouffard2019chemical}, diffusion from the mantle above \citep{jeanloz1990nature,buffett_seagle_2010} or sub-adiabatic thermal effects  \citep{Pozzo_etal_2012}. Within a strongly stratified layer, the dynamics would be very different to the remainder of the convecting core because spherical radial motion would be suppressed \citep{braginsky1999dynamics, davies2015constraints, cox2019penetration}. In terms of using observations of the changing internal geomagnetic field as a window on the dynamics within the core, the existence of a stratified layer is crucial because motion confined to the stratified layer such as waves may have a pronounced geomagnetic signature, which may be falsely interpreted as emanating from the large-scale dynamo process ongoing beneath.

Observational constraints on the stratified layer are largely from seismology, where analysis of a specific `SmKS' class of waves has revealed a localised decrease in wave velocities in the outermost $100-300 \text{km}$ of the core \citep{helffrich2013causes,lay1990stably,helffrich2010outer}, suggesting that the outermost part of the core has a different density and/or elasticity than the rest of the core. 
However, this evidence is far from conclusive because not all studies agree that a stratified layer is necessary to explain seismic measurements \citep{irving2018seismically}, and there are inherent uncertainties due to the remoteness of the core \citep{alexandrakis_eaton_2010}. 
%
So far, observational geomagnetism has offered equivocal evidence for stratified layers. Time dependent observational models can be explained by simple core flow structures on the core-mantle boundary (CMB) which have either no layer \citep{Holme_2015,Amit_2014} (upwelling at the CMB is permitted), or a strongly stratified layer (in which all radial motion is suppressed), \citep{Lesur_etal_2015}.

A complementary approach to understanding the observational signature of a stratified layer is by numerical simulation of a stratified geodynamo model \citep{nakagawa2011effect}.
Models of outer core dynamics have demonstrated that dynamo action can be sensitive to variations in the assumed background state of a fully convective outer core, and that the presence of stably stratified layers can significantly alter the dynamics and morphology of the resultant magnetic field \citep{glane2018enhanced,christensen2018geodynamo,olson2018outer}.
Hence comparisons between the magnetic fields from stratified models with the geomagnetic field can be used to infer compatibility with the presence of a stratified layer. This has been used to constrain the possible thickness of a stratified layer such that it is consistent with geomagnetic observations. \cite{yan2018sensitivity} find that unstratified dynamo simulations significantly underpredict the octupolar component of the geomagnetic field. Their model endorses the presence of a thin stably stratified layer, as the resultant magnetic field can be rendered Earth-like by the inclusion of 60-130 km layer. However, the results are rather sensitive to both the strength of stratification and layer depth, with a thicker layer of 350 km resulting in an incompatible octupole field. 
Similarly \cite{olson2017dynamo} find that stratified model results compare favorably with the time-averaged geomagnetic field for partial stratification in a thin layer of less than 400 km, but unfavorable for stratification in a thick 1000 km layer beneath the CMB. 
Additionally, in terms of dynamics, \cite{Braginsky_93,Buffett_2014} show that MAC (Magnetic, buoyancy (Archimedean) and Coriolis forces) waves in the {\it hidden ocean} at the top of the core provide a mechanism for the 60 year period oscillations detected in the geomagnetic field \citep{roberts200760}. 
The model of \cite{buffett2016evidence} suggests that MAC waves underneath the CMB are also able to account for a significant part of the fluctuations in length of day (LOD) \citep{gross2001combined,holme2005geomagnetic} through explaining the dipole variation, but are contingent on the existence of a stratified layer at the top of the core with a thickness of at least 100 km. 
However, not all stratified dynamo model results champion this scenario for the Earth. It has been found that the inclusion of a thin stable layer in numerical models can act to destablise the dynamo, through generating a thermal wind which creates a different differential rotation pattern in the core \citep{stanley2008effects}.
Additionally many distinctive features of the geomagnetic field are not reproduced, as strong stratification leads to the disappearance of reverse flux patches and suppression of all non-axisymmetric magnetic field components \citep{mound2019regional,christensen2008models}.

One reason why there is no clear message from existing geodynamo models is perhaps that they all have been run in parameter regimes very far from Earth's core \citep{Roberts_Aurnou_2011}. Two important parameters, the Ekman and Rossby numbers, quantify the ratio of rotational to viscous forces $E \sim 10^{-15}$ and the ratio of inertial to viscous forces $R_o \sim 10^{-7}$ respectively \citep{Christensen_2015}. 
These parameters being so small causes difficulties when attempting to numerically simulate the geodynamo because they lead to small spatial and temporal scales that need to be resolved in any direct numerical simulation, but are extremely computationally expensive to do so. Despite this challenge, numerical models have been used with great success to simulate aspects of the geodynamo, reproducing features such as torsional oscillations \citep{Wicht_Christensen_2010} that are consistent with observational models \citep{gillet2010fast}, geomagnetic jerks \citep{aubert2019geomagnetic} and allowing predictions of the Earth's magnetic field strength \citep{christensen2009energy}.
Recent simulations have been able to probe more Earth-like parameter regimes than previously possible, achieving very low Ekman numbers of $E = 10^{-7} - 10^{-8}$ \citep{schaeffer2017turbulent,aubert2019approaching}. However despite this progress, these simulations remain in parameter regimes vastly different to that of the Earth \citep{Christensen_2015}, posing the inescapable question of how representative of the Earth they really are, as force balances can still vary significantly between the simulation regime and the correct regime of the Earth \citep{wicht2019advances}, with the ability to simultaneously reproduce Earth-like field morphology and reversal frequency still beyond current capabilities \citep{christensen2010conditions}. The assessments conducted by \cite{sprain2019assessment} highlight that present geodynamo models able unable to satisfactorily reproduce all aspects of Earth's long term field behaviour.



In this paper we consider the approach proposed by \cite{Taylor_63}, based on the assumption that the inertia-free and viscosity-free asymptotic limit is more faithful to Earth's dynamo than adopting numerically-expedient but nevertheless inflated parameter values. This amounts to setting the values of $R_o$ and $E$ to zero, which simplifies the governing equations significantly, enabling numerical solutions at less computational expense and importantly for us, analytic progress to be made.
The resulting dimensionless magnetostrophic regime then involves an exact balance between the Coriolis force, pressure, buoyancy and the Lorentz force associated with the magnetic field $\bB$ itself:
\begin{equation}\vec{\hat{z}} \times \vec{u} = -\grad p + F_B\vec{\hat r} + (\curl \vec{B}) \times \vec{B}, \label{eqn:magneto}
\end{equation}
where $F_B$ is a buoyancy term that acts in the unit radial direction $\vec{\hat r}$ \citep{Fearn_98}. 



Throughout this paper we consider the magnetostrophic balance of \cref{eqn:magneto}.
\citet{Taylor_63} showed that, as a consequence of this magnetostrophic balance, the magnetic field must obey at all times $t$ the well-known condition
\begin{equation} 
T(s,t) \equiv \int_{C(s)} ((\curl \bB) \times \bB)_\phi ~ s  \text{d}\phi \text{d}z =0,\label{eqn:Taylor} 
\end{equation}
for any geostrophic cylinder $C(s)$ of radius $s$, aligned with the rotation axis, where $(s,\phi,z)$ are cylindrical coordinates.
This constraint applies in the general case for fluids independent of stratification. It was first shown by \cite{malkus1979dynamo} how \eqref{eqn:Taylor} can be refined within a stratified layer of constant depth, which in the limit of zero radial flow leads to a more strict constraint. This constraint now applies on every axisymmetric ring coaxial with the rotation axis that lies within the layer and is known as the {\it{Malkus constraint}}
$$ M(s,z,t) \equiv \int_0^{2\pi} ((\grad \times \vec{B}) \times \vec{B})_{\phi} ~ \text{d}\phi =0, $$
for any $s$ and $z$ within the layer. Magnetic fields that satisfy the Taylor or Malkus constraints respectively are termed Taylor or Malkus states.
%




The associated timescale over which the dominant force balance described by the magnetostrophic equations evolves is $\sim 10^4$ years. 
However observations show changes in the geomagnetic field on much shorter timescale of years to decades \citep{jackson2015geomagnetic}.
This vast discrepancy in timescales motivates distinguishing between the slowly evolving background state and perturbations from it and considering these two features separately.
The theoretically predicted magnetostrophic timescale, represented by Taylor or Malkus states, describes the slow evolution of the magnetic field, and may explain dynamics such as geomagnetic reversals and also the longstanding dominance of the axially symmetric dipolar component of the field. 
Although rapid dynamics such as MHD waves occur on a much shorter timescale, they cannot be considered in isolation as their structure depends critically upon the background state that they perturb. Thus although insightful models of perturbations can be based upon simple states  
\citep[e.g.][]{Malkus_67}, ultimately a close fit to the observed geomagnetic field requires accurate knowledge of the background state. It is the search for such a state that is explored in this paper.

Dynamical models of a non-stratified background state, produced by evolving the magnetic field subject to Taylor's constraint, have appeared very recently \citep{Wu_Roberts_2015,roberts2018magnetostrophic,li2018taylor} and are currently restricted to axisymmetry, although the model of \cite{li2018taylor} can be simply extended to a three dimensional system. These models can additionally be used to probe the effect of incorporating inertia driven torsional waves within this framework \citep{roberts2014modified}.




In this paper we adopt a different strategy and explore the use of both the Taylor and Malkus constraints as a tool for analytically constraining instantaneous structures of the magnetic field throughout Earth's core. This method ignores any dynamics and asks simply whether we can find a set of magnetic fields which satisfy the necessary constraints: Taylor's constraint in the interior and Malkus's constraint in the stratified layer, which will provide plausible background geomagnetic states. 
However, constructing Malkus states is a non-trivial task. Firstly we need to establish whether such fields can even exist, and if so how numerous they are, before we are able to construct examples of Malkus states.
Since we are geophysically motivated, we also wish to determine whether such fields can be compatible with geomagnetic observations.

Our task is a challenging one: even finding magnetic fields that exactly satisfy the comparatively simple case of Taylor's constraint has proven to be difficult in the 55 years since the seminal paper of \citet{Taylor_63}, although notable progress has been made in axisymmetry \citep{Hollerbach_Ierley_91,Soward_Jones_83} and 
in 3D \citep{Jault_Cardin_99} subject to 
imposing a specific symmetry. 
Recently, significant progress has been made in this regard by presenting a more general understanding of the mathematical structure of Taylor's constraint in three dimensions \citep{livermore2008structure}. 
This method was implemented by \cite{livermore2009construction} to construct simple, large scale magnetic fields compatible with geomagnetic observations. It is this which provides the foundation for the work presented here.

The remainder of this paper is structured as follows.
In section 2 we present a new, more general derivation of the condition required to be satisfied with a stratified layer of fluid, which under an idealised limit reduces to what is known as Malkus' constraint. 
In section 3 we summarise the method for discretising and constructing a Taylor state before extending this to Malkus states in section 4. In section 5 we prove that an arbitrary poloidal field can be transformed into a Malkus state through the addition of an appropriate toroidal field and show how this is a useful approach due to the resultant equations being linear. In section 6 we present our results for an Earth like magnetic field satisfying all relevant constraints, within the linear framework. In section 7 we discuss these results with regard to Earth's internal field, specifically our estimate of toroidal field strength, before 
concluding in section 8.

%
%

\section{Derivation of Malkus' constraint} \label{sec:Malk_derivation}

Within stably stratified fluids radial flows are suppressed, hence in the limit of strong stratification radial fluid velocities are negligibly small \citep{braginsky1999dynamics,davies2015constraints}. 
We proceed within this idealistic limit and require that $u_r=0$ within a region of stratified fluid that is a volume of revolution: we represent the proposed stratified layer within Earth's core as a spherically symmetric layer of constant depth. We assume further that the system is in magnetostrophic balance; that is, rapidly rotating with negligible inertia and viscosity. The resulting constraint was first derived by \cite{malkus1979dynamo}, however, here we present an alternative and more straightforward derivation courtesy of Dominique Jault (personal communication).

\begin{figure}[H]
	\centering
	\begin{subfigure}{.48\textwidth}
		\centering
		\includegraphics[width=0.55\textwidth]{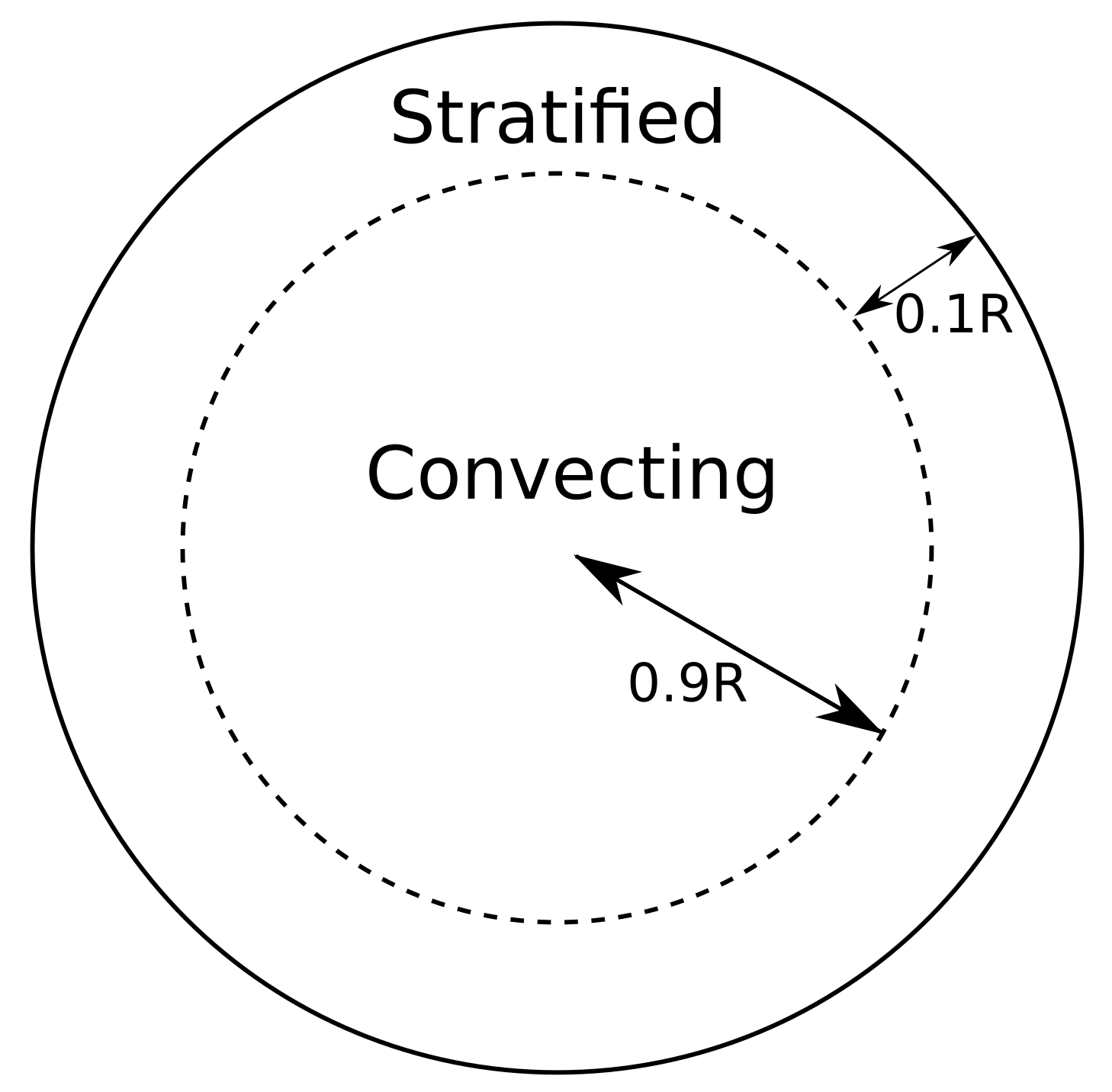}
 \caption{Earth-like spherical shell with radius \\$r_{SL}=0.9R$. A Malkus state defined in a stratified layer surrounds an interior Taylor state. \label{fig:fulldomain}}
		
	\end{subfigure}%
	\begin{subfigure}{.48\textwidth}
	\centering
\includegraphics[width=0.98\textwidth]{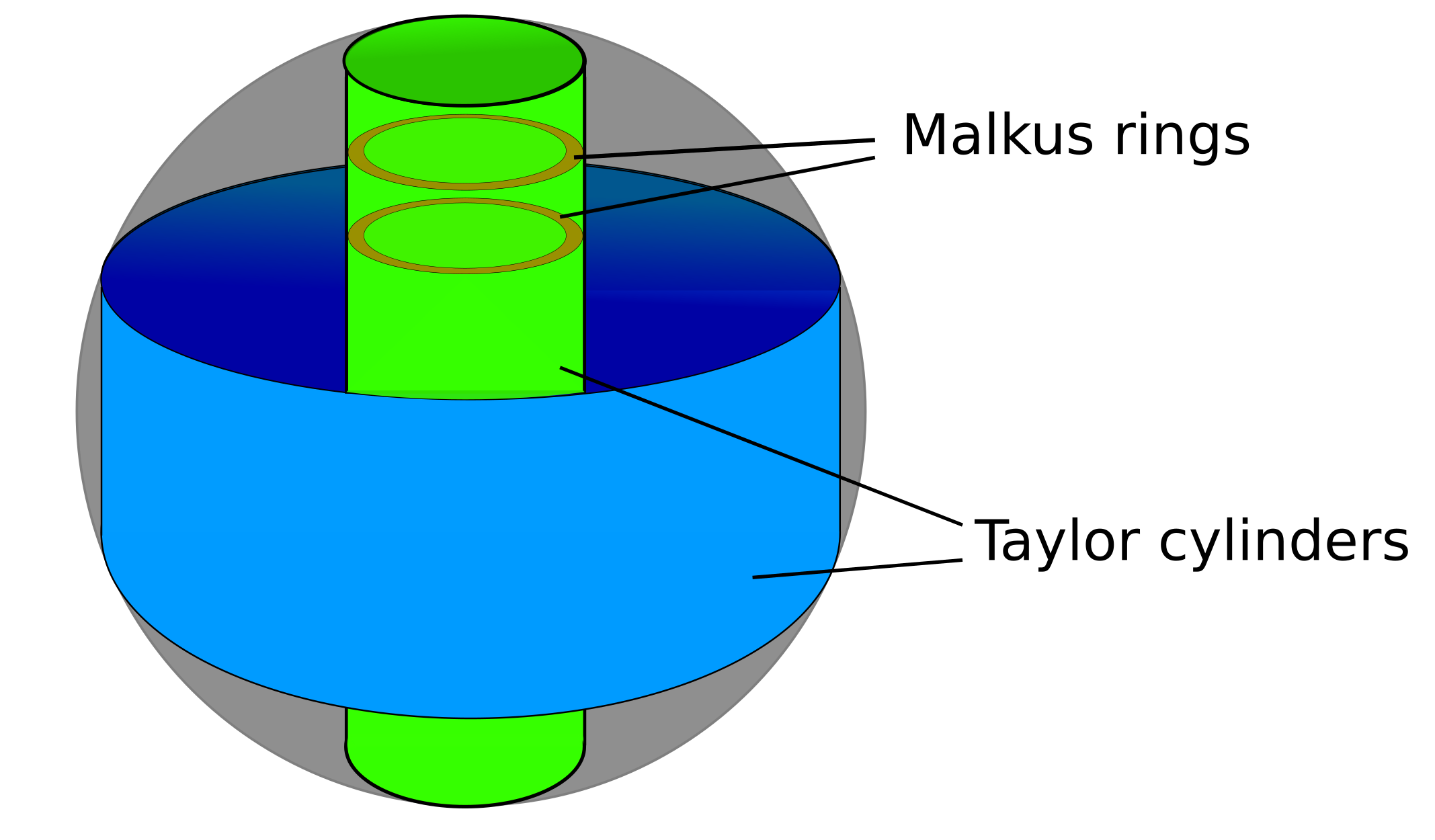}
 \caption{ Geometry of constraint surfaces \label{fig:constraint_surfaces}}
 \end{subfigure}
  \end{figure}

We use the condition for incompressible flow that $\grad \cdot \vec{u} = 0$ and the standard toroidal poloidal decomposition within spherical coordinates $(r,\theta,\phi)$. 
From the condition that there is no spherically-radial component of velocity then $\vec{u}$ must be purely toroidal and hence can be written as $$\vec{u} = \grad \times (\mathcal{T}(r,\theta,\phi)\vec{\hat{r}})= \frac{1}{r\sin\theta} \pd{\mathcal{T}}{\phi} \vec{\hat{\theta}} - \frac{1}{r} \pd{\mathcal{T}}{\theta} \vec{\hat{\phi}}.$$
Therefore the cylindrically-radial velocity, written in spherical coordinates, is
$$u_s=\sin\theta u_r+\cos\theta u_\theta 
= \frac{\cos\theta}{r\sin\theta} \pd{\mathcal{T}}{\phi} $$
and so
$$\int_0^{2\pi} u_s ~ \text{d}\phi = \frac{\cos\theta}{r\sin\theta} \int_0^{2\pi} \pd{\mathcal{T}}{\phi} ~ \text{d}\phi = 0. $$
%
%
Now, since $\vec{\hat{\phi}} \cdot (\zhat \times \vec{u})  = u_s $ then, from the azimuthal component of the magnetostrophic \cref{eqn:magneto} we have
$$u_s = -\pd{p}{\phi} +((\grad \times \vec{B}) \times \vec{B})_{\phi}.$$
Integrating this around any circle in a plane orthogonal to $\vec{\hat{r}}$ centred on the rotation axis, (as illustrated by the red rings in \cref{fig:constraint_surfaces}), and using the single-valued nature of pressure, gives Malkus' constraint,

\begin{equation} \underbrace{\int_0^{2\pi}u_s ~ \text{d}\phi}_{=0} = -\underbrace{\int_0^{2\pi}\pd{p}{\phi} ~ \text{d}\phi}_{=0} +\int_0^{2\pi} ((\grad \times \vec{B}) \times \vec{B})_{\phi} ~d\phi = 0, \nonumber  \end{equation}
or equivalently requiring that the Malkus integral $M$ is zero:
\begin{equation}
    M(s,z,t) \equiv \int_0^{2\pi} ((\grad \times \vec{B}) \times \vec{B})_{\phi} ~ \text{d}\phi =0. \label{eqn:Malkcon}
\end{equation}

We are also able to generalise this constraint from considering the idealistic limit of requiring $u_r=0$ within the stratified fluid to the more general situation of permitting $u_r \neq 0$, where we express the Malkus integral in terms of the radial flow. Now, the flow $\bu$ is no longer purely toroidal and hence
\begin{eqnarray} M(s,z,t) 
=\int_0^{2\pi} u_s ~ \text{d}\phi = \int_0^{2\pi} u_\theta \cos\theta \text{d}\phi + \int_0^{2\pi} u_r \sin\theta ~ \text{d}\phi. \end{eqnarray}



We now use the condition for incompressible flow that $\grad \cdot \vec{u} = 0,$
$$ 0 = \grad \cdot \bu = \frac{1}{r^2}\pd{(r^2u_r)}{r} +\frac{1}{r\sin\theta}\pd{(u_\theta \sin\theta)}{\theta} + \frac{1}{r\sin\theta} \pd{u_\phi}{\phi},$$

$$ \Rightarrow \int_0^{2\pi} \left(\frac{\sin\theta}{r}\pd{(r^2u_r)}{r} + \pd{(u_\theta \sin\theta)}{\theta} \right) \text{d}\phi = - \int_0^{2\pi} \pd{u_\phi}{\phi} \text{d}\phi = 0.$$
Now integrating over $[0, \theta]$ we find
$$ \int_0^{2\pi}u_\theta \text{d}\phi = \frac{1}{\sin\theta}\int_0^\theta \frac{\sin\theta'}{r} \int_0^{2\pi}\pd{(r^2u_r)}{r} \text{d}\phi \text{d}\theta' = - \frac{1}{r\sin\theta}\int_0^\theta \sin \theta' \pd{}{r}\left( r^2 \int_0^{2\pi} u_r d\phi\right) ~ \text{d}\theta' $$

$$\Rightarrow M = -\frac{1}{r\tan\theta} \int_0^\theta \pd{}{r} \left(r^2\int_0^{2\pi} u_r \sin\theta' \text{d}\phi \right) \text{d}\theta' + \int_0^{2\pi} u_r \sin\theta ~ \text{d}\phi.$$


%
In the above derivation, no assumption has been made about stratification and this equation holds as an identity in the magnetostrophic regime independent of stratification.
In the case considered by Malkus, $M=0$ is recovered in the limit of $u_r \rightarrow 0$.

It is clear that Malkus' constraint is similar to Taylor's constraint except now not only does the azimuthal component of the Lorentz force need to have zero average over fluid cylinders, it needs to be zero for the infinite set of constant-$z$ slices of these cylinders (here termed {\it Malkus rings}, see figure \ref{fig:constraint_surfaces}) that lie within the stratified region. 
%
In terms of the flow, the increased restriction of the Malkus constraint arises because it requires zero
azimuthally-averaged $u_s$ at any given value of $z$, whereas Taylor's constraint requires only that the cylindrically averaged $u_s$ vanishes and allows outward flow at a given height to be compensated by inward flow at another. 
We note that all Malkus states are Taylor states, but the converse is not true.

  \section{Geometry and representation of a stratified magnetostrophic model} \label{sec:fulldom}
  
  The physical motivation for applying Malkus' constraint arises from seeking to represent a realistic model for the magnetic field in the proposed stratified layer within Earth's outer core. 
   Hence we compute solutions for the magnetic field in the  Earth-like configuration illustrated in \cref{fig:fulldomain}, consisting of a spherical region in which Taylor's constraint applies, representing the convective region of Earth's core, surrounded by a spherical shell in which Malkus' constraint applies, representing the stratified layer immediately beneath the CMB.
   Our method allows a free choice of inner radius $r_{SL}$, so in order to agree with the bulk of seismic evidence \citep{helffrich2010outer,helffrich2013causes,lay1990stably}, the value $r_{SL}=0.9R$ is chosen for the majority of our solutions, where $R$ is the full radius of the core (3845 km).
   However due to the uncertainty which exists for the thickness of Earth's stratified layer \citep{Kaneshima_2017}, we also probe how sensitive our results are to layer thickness, considering $r_{SL}=0.85R$ and $r_{SL}=0.95R$ as well.
   The Earth's inner core is neglected throughout, since incorporating it would lead to additional intricacies due to the cylindrical nature of Taylor's constraint which leads to a distinction between regions inside and outside the tangent cylinder \citep{Livermore_Hollerbach_2012,livermore2008structure}. Since the focus here is on the outermost reaches of the core, we avoid such complications.
   
      The method used to construct the total solution for the magnetic field throughout Earth's core that is consistent with the Taylor and Malkus constraints is sequential. Firstly, we use a regular representation of the form shown in \cref{eqn:torpolexpan} to construct a Malkus state in the stratified layer. Secondly, we construct a Taylor state which matches to the Malkus state at $r=r_{SL}$; overall the magnetic field is continuous but may have discontinuous derivatives on $r=r_{SL}$. We note that any flow driven by this magnetic field through the magnetostrophic balance may also be discontinuous at $r=r_{SL}$ because in general $u_r \neq 0$ in the inner region but $u_r = 0$ is assumed in the stratified region. 
      Considerations of such dynamics lie outside the scope of the present study focussed only on the magnetic constraints, but imposing continuity of $u_r$ for example would clearly require additional constraints. 

   As a pedogogical exercise we also construct some Malkus states within a fully stratified sphere ($r_{SL} = 0$), as detailed in \cref{sec:Apa_both}. Without the complications of matching to a Taylor state, the equations take a simpler form and we present some first examples in \cref{sec:Ap_sol_simp}.
   Dynamically, sustenance of a magnetic field within a fully stratified sphere is of course ruled out by the theory of \cite{Busse_75a}, which provides a strictly positive lower bound for the radial flow as a condition on the existence of a dynamo.
%
 %
 Nonetheless it can be insightful to first consider the full sphere case, as it facilitates the consideration of fundamental principles of the magnetic field and Malkus constraint structure, and allows direct comparisons to be made with similar full sphere Taylor states.

In what follows we represent a magnetic field by a sum of toroidal and poloidal modes with specific coefficients
\begin{equation} \label{eqn:Brep} \vec{B} = \sum_{l=1}^{L_{max}} \sum_{m=-l}^{l} \sum_{n=1}^{N_{max}} a_{l,n}^m {\vec{\mathcal{T}} }_{l,n}^m + b_{l,n}^m {\vec{\mathcal{S}}}_{l,n}^m \end{equation}
where
$\vec{\mathcal{T}}_{l,n}^m=\curl (T_{l,n}(r) Y_l^m \rhat)$, $\vec{\mathcal{S}}_{l,n}^m=\curl \curl (S_{l,n}(r) Y_l^m \rhat)$, $N_{max}$ is the radial truncation of the poloidal and toroidal field. In the above, $Y_l^m$ is a spherical harmonic of degree $l$ and order $m$, normalised to unity by its squared integral over solid angle. Positive or negative values of $m$ indicate respectively a $\cos m\phi$ or $\sin m\phi$ dependence in azimuth. The scalar functions ${{T} }_{l,n}^m$ and ${{S} }_{l,n}^m$, $n\ge 1$, are respectively chosen to be the functions $\chi_{l,n}$ and  $\psi_{l,n}$ composed of Jacobi polynomials  \citep{li2010optimal,Li_etal_2011}. They are orthogonal, and obey regularity conditions at the origin and the electrically insulating boundary condition at $r=R$
\begin{equation} \frac{d \mathcal{S}_l^m}{dr} + l \mathcal{S}_l^m/R = \mathcal{T}_l^m = 0. \label{eqn:bc}
\end{equation}
We note that this description is convenient but incomplete when used within the spherical shell, for which the magnetic field does not need to obey regularity at the origin. For simplicity, we nevertheless use this representation in both layers, although restricting the domain of the radial representation to $[0,r_{SL}]$ for the inner region.
  
  
 %

   

\section{Discretisation of the Taylor constraint} \label{sec:Tay_disc}



%
Since the Malkus constraint forms a more restrictive constraint which encompasses the Taylor constraint it is useful for us to first summarise the structure of the Taylor constraint in a full sphere.
The integral given in \cref{eqn:Taylor}, which Taylor's constraint requires to be zero, is known as the Taylor integral.
Although applied on an infinite set of surfaces, 
\cite{livermore2008structure} showed that Taylor's constraint reduces to a finite number of constraint equations for a suitably truncated magnetic field expansion
\begin{equation} \mathcal{S}_l^m(r)=r^{l+1}\sum_{j=0}^{N_{max}}c_j r^{2j} ~~~~ \text{and} ~~~~ \mathcal{T}_l^m(r)=r^{l+1}\sum_{j=0}^{N_{max}}d_j r^{2j}, \label{eqn:torpolexpan} \end{equation}
which is an expanded version of \eqref{eqn:Brep} for some $c_j$ and $d_j$. The Taylor integral itself then collapses to a polynomial of finite degree 
%
%
%
%
%
which depends upon $s^2$ \citep{lewis1990physical} and the coefficients $a_{l,n}^m, b_{l,n}^m$, and takes the form

\begin{equation} \label{eqn:Taypoly} T(s) = s^2\sqrt{R-s^2}Q_{D_{T}}(s^2)=0,
\end{equation}
for some polynomial $Q_{D_{T}}$ of maximum degree $D_T$. 

Taylor's constraint is now equivalent to enforcing that the coefficients of all powers of $s$ in the polynomial $Q_{D_{T}}$ equal zero, as this ensures $T(s)$ vanishes identically on every geostrophic contour. This reduces the infinite number of constraints to a finite number of simultaneous, coupled, quadratic, homogeneous equations.  %
This reduction is vital as it gives a procedure for enforcing Taylor's constraint in general, and allows the implementation of a method to construct magnetic fields which exactly satisfy this constraint, known as Taylor states, as demonstrated by \cite{livermore2009construction}. 
In the next section we see how, with some relatively simple alterations this procedure can be extended to the construction of exact Malkus states. 

\section{Malkus states} \label{sec:Malk_state}

This section outlines some general properties of the mathematical structure of Malkus' constraints and provides the methodology for constructing the first known Malkus states; we also address the questions of existence and uniqueness of solutions and the dimension of the resultant solution space.

Along similar lines as we showed for Taylor's constraints in \cref{sec:Tay_disc}, on adopting the representation \eqref{eqn:Brep} the Malkus integral reduces to a multinomial in $s^2$ and $z$ \citep{lewis1990physical} 
and we require
$$M(s,z) = Q_{D_{M}}(s^2,z) = 0$$
for some finite degree multinomial $Q_{D_{M}}$ in $s$ and $z$. Note that the Taylor integral \eqref{eqn:Taypoly} is simply a z-integrated form of $Q_{D_{M}}$. 
Equating every multinomial term in $Q_{D_{M}}(s^2,z)$ to zero results in a
finite set of constraints that are nonlinear in the coefficients $a_{l,n}^m$ and $b_{l,n}^m$.

The number of constraints can be quantified for a given truncation following a similar approach as that employed by \cite{livermore2008structure} for Taylor's constraint, by tracking the greatest exponent of the dimension of length. This analysis is conducted in \cref{sec:enum_con} and results in the number of Malkus constraints given by 
\begin{equation} \label{eq:Malk_numcon}
    C_M= {C_T}^2+3C_T+2,
\end{equation}
where the number of Taylor constraints for an equivalent magnetic field is $C_T = L_{max} + 2N_{max} - 2$ (after the single degeneracy due to the electrically insulating boundary condition is removed) \citep{livermore2008structure}.
Therefore we find that as expected the Malkus' constraints are more numerous than Taylor's constraints. 
It is significant to notice that $C_M \gg C_T$ and in particular for high degree/resolution systems $C_M \approx {C_T}^2$.

In order to satisfy these constraints, the magnetic field has $2L_{max}N_{max}(L_{max}+2)$ degrees of freedom (this being the number of unknown spectral coefficients within the truncation of $(L_{max}, N_{max})$. 
In axisymmetry the number of degrees of freedom reduces to $2N_{max}L_{max}$. 



If we truncate the magnetic field quasi uniformly as $N= \mathcal{O}(L_{max}) \approx \mathcal{O}(N_{max})$, then we observe that at high $N$ the number of constraints ($O(N^2)$ Malkus constraints; $O(N)$ Taylor constraints) is exceeded by the number of degrees of freedom of $N^3$. 
A simple argument based on linear algebra suggests that many solutions exist at high $N$, however this may be misleading because the constraints are nonlinear and it is not 
obvious \textit{a priori} whether any solutions exist, or if they do, how numerous they might be.

We consider a simple example in \cref{sec:Apb}, which shows the structure of constraint equations that arise. The example highlights that degeneracy of the constraint equations plays a far more significant role for the Malkus constraints compared with the Taylor constraints, which only have a single weak degeneracy due to the electrically insulating boundary condition \citep{livermore2008structure}.
%
%
However, due to the complex nature of these nonlinear equations, at present we have no theory to predict which constraints will be degenerate and hence the total number of independent constraints. 


Because of the apparent uncertainty of the existence of Malkus states, it is instructive to identify  whether imposing strong symmetry is useful to identify very simple examples.
Owing to symmetries inherent in the spherical harmonics, many classes of simple Taylor states exist, as outlined by \cite{livermore2009construction}: for example any field that is either symmetric or anti-symmetric with respect to a rotation of $\pi$ radians about the $x$-axis is a Taylor state. Due to the absence of averaging in $z$, such symmetric magnetic fields do not automatically satisfy the Malkus constraints. However some simple classes of field are guaranteed to be Malkus states, such as single spherical harmonic modes, axisymmetric purely toroidal or poloidal fields since the integrand itself $((\curl \bB) \times \bB)_\phi$ is zero. 
Also equatorially symmetric purely toroidal or poloidal fields comprising either only cosine or only sine dependence in azimuth are Malkus states as the resultant integrand is anti-symmetric with respect to a rotation of $\pi$ radians and hence the azimuthal average over $[0,2\pi]$ causes the Malkus integral to vanish.

\section{Finding a Malkus state}
Owing to the nonlinear albeit finite nature of the Malkus constraints, it is far from obvious whether any solutions exist beyond those of the simple structure explored above. In the next section, we demonstrate the existence of a class of solutions with arbitrarily complex lateral structure.

\subsection{A special class of Malkus states} \label{sec:theo}

Here we demonstrate that within the class of magnetic fields that all contain a known poloidal component (but whose toroidal component is unknown) then there exists systems where all the Malkus constraints are linear in the unknown spectral parameters. A formal statement of this fact is given in the theorem given below.  

\begin{theorem}
Any arbitrary, prescribed, polynomial poloidal field can be transformed into a Malkus state through the addition of an appropriate polynomial toroidal field.
\end{theorem}



\begin{figure}[H] 
	\centering
	\includegraphics[width=0.7\textwidth]{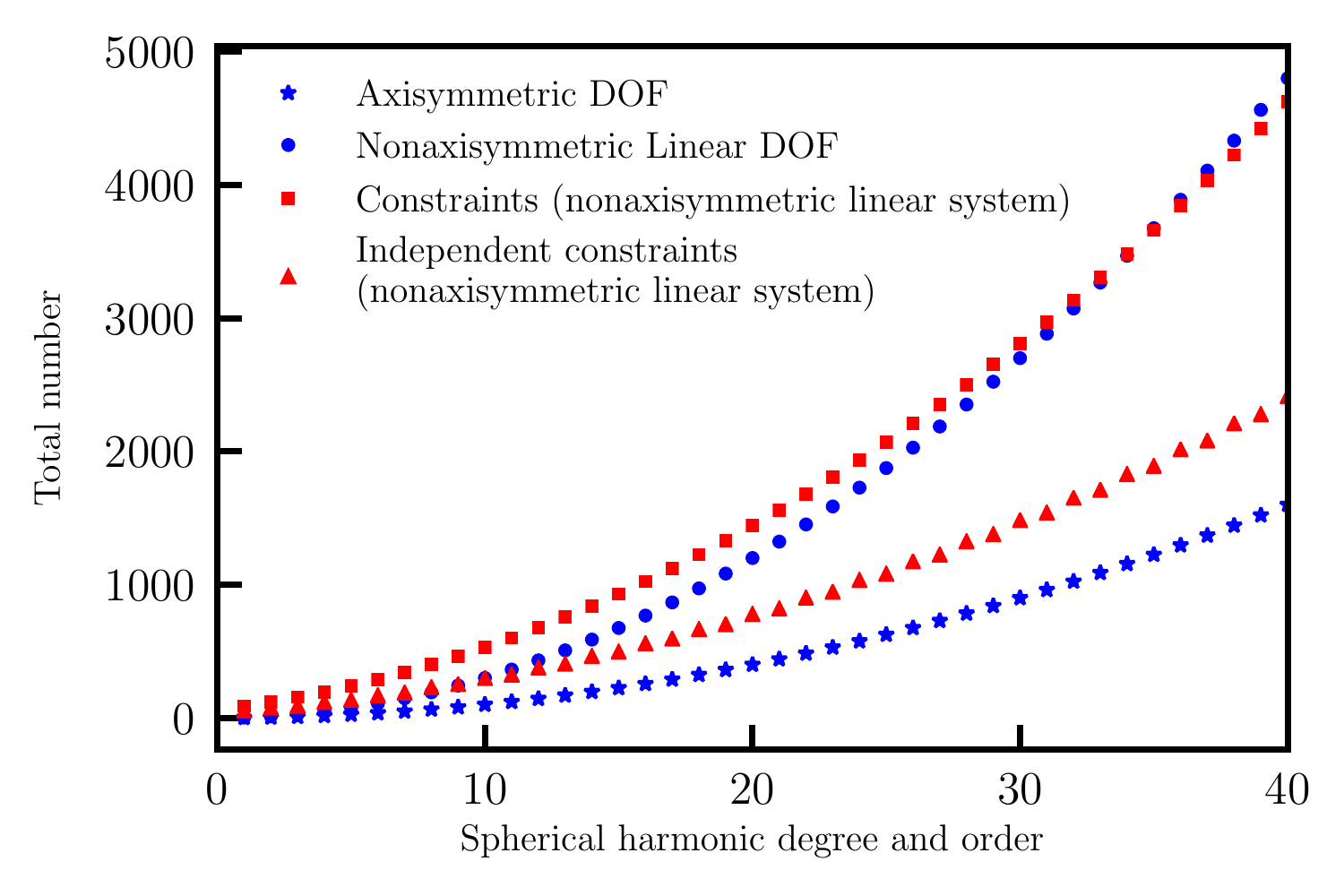}
 \caption{This graph compares the number of constraints to degrees of freedom (DOF) as a function of toroidal field spherical harmonic resolution with $L_{max}=M_{max}=N_{max}$, given a fixed poloidal field of $L_{max}=M_{max}=13$. This illustrates that for the non-axisymmetric linear system we construct then the number of degrees of freedom (red) exceeds the number of independent constraints (red triangles) for a toroidal field of resolution $L_{max}=N_{max} \geq 10$. \label{fig:new_con_dof}}

\end{figure}

\begin{proof} \label{proof:linear}
We prove below that by considering an arbitrary, prescribed, truncated polynomial poloidal field, the addition of a specific choice of toroidal modes renders the Malkus constraints linear in the unknown toroidal coefficients. By taking a sufficient number of such modes such that the degrees of freedom exceed the number of constraints, it follows that for the general case (barring specific degenerate cases) by solving the linear system the resultant magnetic field is a Malkus state.

To show this, because the Malkus constraint is quadratic in the magnetic field, we introduce the concept of a magnetic field interaction. In general there are three possible field interactions within the Malkus integral, toroidal-toroidal, poloidal-poloidal and toroidal-poloidal, respectively
$$M = \sum_{l_1,l_2}^{L_{max}} \sum_{m}^{L_{max}} \left( [\vec{T}_{l_l}^m, \vec{T}_{l_2}^m] + [\vec{S}_{l_l}^m, \vec{S}_{l_2}^m] + [\vec{T}_{l_l}^m, \vec{S}_{l_2}^m] \right)$$
where

 \begin{align}
 \label{eq:tortor}
    [\vec{T}_{l_l}^m, \vec{T}_{l_2}^m] &= \int_0^{2\pi} \frac{l_1(l_1+1)\mathcal{T}_{l_l}^m \mathcal{T}_{l_2}^m}{r^3 \sin\theta}\left({Y}_{l_l}^m\pd{{Y}_{l_2}^m}{\phi}\right) s ~ \text{d}\phi +sc, \\
    [\vec{S}_{l_l}^m, \vec{S}_{l_2}^m] &= \int_0^{2\pi} \frac{l_1(l_1+1)\mathcal{S}_{l_l}^m (\frac{\text{d}^2}{\text{d}r^2}-l_2(l_2+1)/r^2)\mathcal{S}_{l_2}^m}{r^3 \sin\theta}\left({Y}_{l_l}^m\pd{{Y}_{l_2}^m}{\phi}\right) s ~ \text{d}\phi +sc,\nonumber \\
    [\vec{T}_{l_l}^m, \vec{S}_{l_2}^m] &= \int_0^{2\pi} \frac{1}{r^3}\left( l_1(l_1+1){T}_{l_l}^m \frac{\text{d}\mathcal{S}_{l_2}^m}{\text{d}r} Y_{l_1}^m \pd{{Y}_{l_2}^m}{\theta}\right.  
    \left.- l_2(l_2+1)\mathcal{S}_{l_2}^m \frac{\text{d}T_{l_1}^m}{\text{d}r} Y_{l_2}^m \pd{{Y}_{l_1}^m}{\theta} \right) s ~ \text{d}\phi, \nonumber 
 \end{align}
where $sc$ is the symmetric counterpart given by interchanging the vector harmonics and hence the positions of $l_1$ and $l_2$ \citep{livermore2008structure}. Note that there is no poloidal-toroidal interaction since the curl of a poloidal vector is toroidal and ($\vec{\mathcal{T}_1} \times \vec{\mathcal{T}_2})_\phi = 0,$ for any two toroidal vectors $\vec{\mathcal{T}_1}$ and $\vec{\mathcal{T}_2}$.

For the situation we consider of a given poloidal field, then the only non-linearity within the unspecified coefficients arises from the toroidal-toroidal interactions, which results in quadratic dependence, just as for the general case with unprescribed poloidal field. 
However, by restricting attention to toroidal fields that result in no toroidal-toroidal interaction, the unknown toroidal coefficients appear only in a linear form through the toroidal-poloidal interactions. 
Axisymmetric modes are the simplest set of toroidal modes which are non-self-interacting, however there are too few of them (within the truncation) to solve the resulting linear system which is  over-constrained (see \cref{fig:new_con_dof}).

Therefore we require additional non-axisymmetric toroidal modes, which we choose such that the total set of toroidal modes remains non-self-interacting. This is achieved by exploiting the previously noted observations that any single harmonic is a Malkus state and that the set of equatorially symmetric toroidal modes $T_l^l$ is a Malkus state (and therefore has no self-interaction). Owing additionally to azimuthal symmetry, the modes 
$${T_1^0}, T_2^0, \cdots, T_1^{-1}, T_{1}^1,T_2^{-2}, T_{2}^2, \dots, $$
that is, the modes $T^m_l$ with $m = 0$ or $m = \pm l$, have no self-interactions. Each harmonic mode may be expanded in radial modes up to the truncation $N_{max}$. The non-interacting nature of the modes may be confirmed from \cref{eq:tortor}.

%
%

The addition of these nonaxisymmetric modes increases the number of degrees of freedom from the axisymmetric case by a factor of three such that it is now larger than the number of constraints (which are now all linear).
This can be shown in general since for a toroidal field truncated at $L_1, N_1$ and a poloidal field truncated at $L_2, N_2$ the number of Taylor constraints is equal to half of the maximum degree of the polynomial in $s$, (i.e. $C_T = \frac{1}{2}(L_1+L_2+2N_1+2N_2) - 2$) \citep{livermore2008structure} and the maximum number of Malkus constraints we have shown is given in terms of $C_T$ by \cref{eq:Malk_numcon}. This results in a situation where if the poloidal field is fixed at a chosen resolution then for a toroidal field truncated quasi uniformly as $N= \mathcal{O}(L_{max}) \approx \mathcal{O}(N_{max})$ then we can see that the number of Malkus constraints scales as $\frac{9}{4} N^2$, which importantly, grows slower than the number of degrees of freedom for the non-axisymmetric linear system which scales as $3 N^2$. Hence it is guaranteed that at a sufficiently large resolution toroidal field representation then there will be more degrees of freedom than constraints.

Therefore, barring degenerate cases, Malkus states exist. 
Compared with the case of a purely axisymmetric toroidal field, the number (but not the specific form) of linear constraints remains unaltered by the addition of these extra non-axisymmetric modes. 

It is worth noting that the depth of the stratified layer does not enter into above derivation.
The magnetic field solution in fact satisfies the Malkus constraints everywhere within its region of definition: in our case, this is the full sphere $0 \le r \le R$. 
\end{proof}


\Cref{fig:new_con_dof} provides a specific example of the number of constraints given a poloidal field of degree $13$. It demonstrates two important things. Firstly, that due to degeneracy (for which we have no explanation) the independent linear constraints (red triangles) are much fewer than the full set of linear constraints (red squares). Secondly, that the number of degrees of freedom exceed the number of independent constraints at $L_{max}=N_{max} \geq 10$ if we consider the non-axisymmetric toroidal basis (blue circles) but is not exceeded at any truncation if we adopt the axisymmetric toroidal basis (blue stars). In particular, taking a non-axisymmetric toroidal field with truncation $L_{max}=N_{max}=13$ gives an infinite set of Malkus states. 

We note that the above deviation is based upon a polynomial representation, which is sufficient for our purposes here. However, we know that any continuous function defined on a closed interval can be uniformly approximated as closely as desired by a polynomial function, and hence it can be extended to include an arbitrary magnetic field structure by expressing the relevant scalars in a polynomial basis of suitably large truncation.

We need to match the Malkus state (physically defined within the stratified layer) to a Taylor state in the region beneath. One way of proceeding is to simply evaluate the Malkus state beneath the stratified layer (where it also satisfies Taylor constraint); however this effectively imposes additional constraints on the inner region and is overly restrictive. Instead, we impose the same profile of poloidal field and expand the toroidal component of the Taylor state in the same set of spherical harmonic modes as used for the Malkus state. Such a choice also renders the Taylor constraints linear in the unknown toroidal coefficients.

\subsection{Further geophysical constraints}

In order to construct a Malkus state according to the above procedure, we need to completely specify the poloidal field. 
Following \cite{livermore2009construction}, we downwards continue observation-derived models inside the core $r \le R$ by assuming a profile for each poloidal harmonic of degree $l$ that minimises the Ohmic dissipation within the modelled core
\begin{equation} \label{eqn:polprofile}
    (2l+3)r^{l+1} - (2l+1)r^{l+3}.
\end{equation} 
%

We adopt two choices of observation-derived model. First, we use the CHAOS-6 model \citep{Finlay_etal_2016} at epoch 2015 evaluated to degree 13, 
the maximum obtainable from geomagnetic observations without significant interference due to crust magnetism \citep{Kono2015Geointro}. Second, we use the time-averaged field over the last 10000 years from the CALS10k.2 model \citep{constable2016persistent}, which although is defined to degree 10 it has power concentrated mostly at degrees 1--4 because of strong regularisation of sparsely-observed ancient magnetic field structures. Recalling that the magnetostrophic state that we seek is defined over millenial timescales, this longer average provides on the one hand a better approximation to the background state, but on the other a much lower resolution. 

Even within these geomagnetically consistent Malkus states, there are nevertheless multiple degrees of freedom remaining. This raises the question of which of the multiple possible solutions are most realistic for the Earth, and motivates us to incorporate additional conditions to distinguish `Earth-like' solutions. 


We determine specific solutions through optimising the toroidal field $\bf T$ through either its Ohmic dissipation or its energy, respectively
$$Q = \frac{\eta}{\mu_0}\int_V (\curl {\bf T})^2 dV, \qquad \mathcal{E} = \frac{1}{2\mu_0} \int_V {\bf T}^2 dV,$$
where $\eta \approx 1$~m$^2$s$^{-1}$ is magnetic diffusivity and $\mu_0=4\pi \times 10^{-7}~\text{NA}^{-2}$ is the permeability of free space. Both of these target functions are quadratic in the magnetic field, and so seeking a minimal value subject to the now linear  constraints is straightfoward. In our sequential method to find a matched Malkus-Taylor state, we first optimise the Malkus state, and then subsequently find an optimal matching Taylor state.

Of the dissipation mechanisms in the core: Ohmic, thermal and viscous, the Ohmic losses are believed to dominate. 
%
On these grounds, the most efficient arrangement of the geomagnetic field would be such that Ohmic dissipation $Q$ is minimised. 
It is worth noting that 
in general our procedure is not guaranteed to provide the Malkus state field with least dissipation, but only an approximation to it, since we effectively separately optimise for the poloidal and toroidal component with least dissipation.
In terms of finding a Malkus state with minimum toroidal field energy, this is useful in allowing us to determine the weakest toroidal field which is required in order to transform the imposed poloidal field into a Malkus state.
%

%
%


In \cref{sec:Ap_sol_simp} we compare the method of finding the weakest toroidal field required to make a Malkus state, between using only selected toroidal modes, and all toroidal modes (resulting in a nonlinear system). For low truncation, minimisation of the toroidal energy subject to these nonlinear constraints is computationally solvable, and the two approaches produce comparable results.
This suggests that estimates for the lower bound of Earth's toroidal field strength obtained using our linearised approach will not differ greatly from related full non-linear optimisation (that is computationally infeasible).


\section{An Earth-like example} \label{sec:highres}

We now present some visualisations of the specific class of Malkus states discussed above with minimal toroidal field energy
for which the system of equations which enforce the constraints is linear. 
The geometry assumed here is as illustrated in \cref{fig:fulldomain}, with a Malkus state in the stratified layer in the region $0.9R < r \leq  R$, matching to an inner Taylor state. 
We shall show the adjustment of the imposed poloidal field structure to a Malkus state by the required additive toroidal field. The strength of this toroidal field will be shown by contour plots of its azimuthal component.
We note that the radial component of the magnetic field is defined everywhere by the imposed poloidal field, with the smooth degree 2 radial profile defined in \cref{eqn:polprofile}.

\subsection{Magnetic field at 2015} \label{sec:Tay_Malk_sol}
\begin{figure}[H] 
	\centering
	\begin{subfigure}{.47\textwidth}
		\centering
	\includegraphics[width=0.9\textwidth]{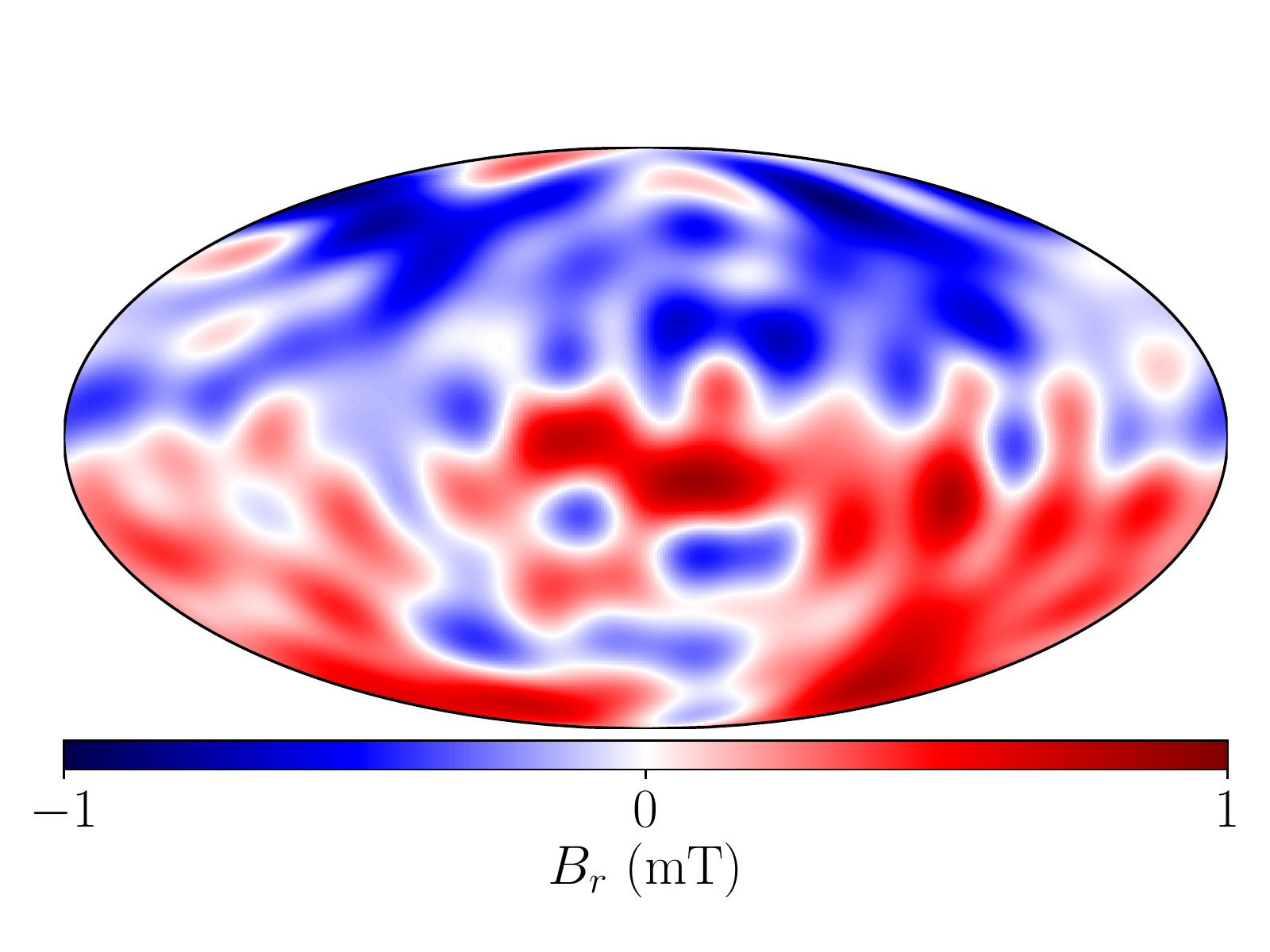}
      \caption{\label{fig:CMB_field_basemap_cor} $B_r$ at the CMB, (max value = 0.91 mT). }
	\end{subfigure}%
	\begin{subfigure}{.47\textwidth}
	\centering
 \includegraphics[width=0.9\textwidth]{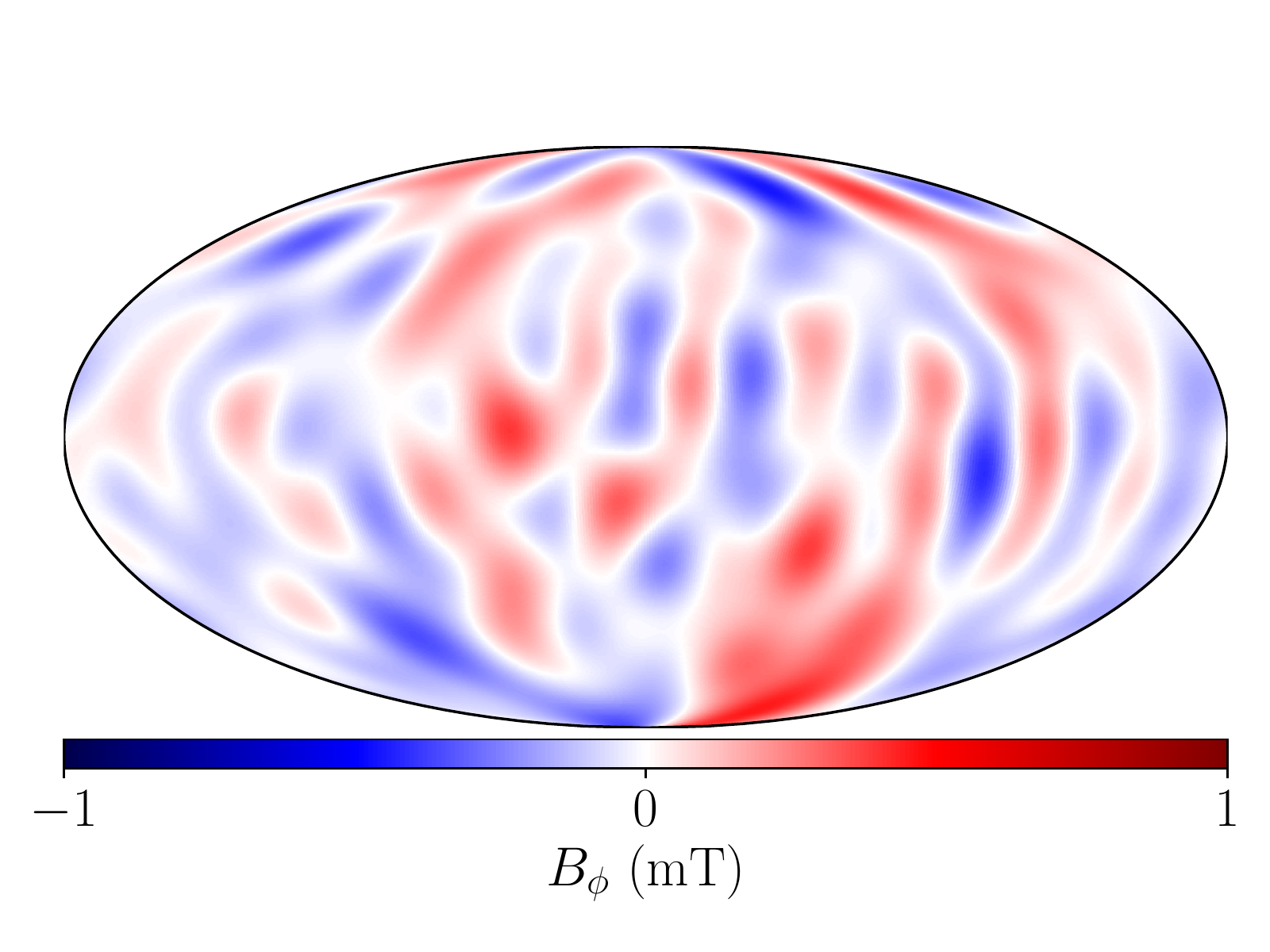}
  \caption{\label{fig:CMB_field_basemap_cor_08} $B_\phi$ at the CMB, (max value = 0.46 mT). }
 \end{subfigure}
 \caption{Magnetic field at the CMB based on the poloidal field from CHAOS-6 at epoch 2015. Visualised using the Mollweide projection and centred on the Greenwich meridian.
 \label{fig:CMB_field_basemap_cor_both}}
   \end{figure}

We begin by showing in \cref{fig:CMB_field_basemap_cor_both} both the radial and azimuthal structure, $B_r$ and $B_\phi$, of the CHAOS-6 model at epoch 2015 on the CMB, $r=R$. Of note is that at the truncation to degree 13, the azimuthal component is about half as strong as the radial component.

\begin{figure}[H]
  \centering
  \includegraphics[width=0.7\textwidth]{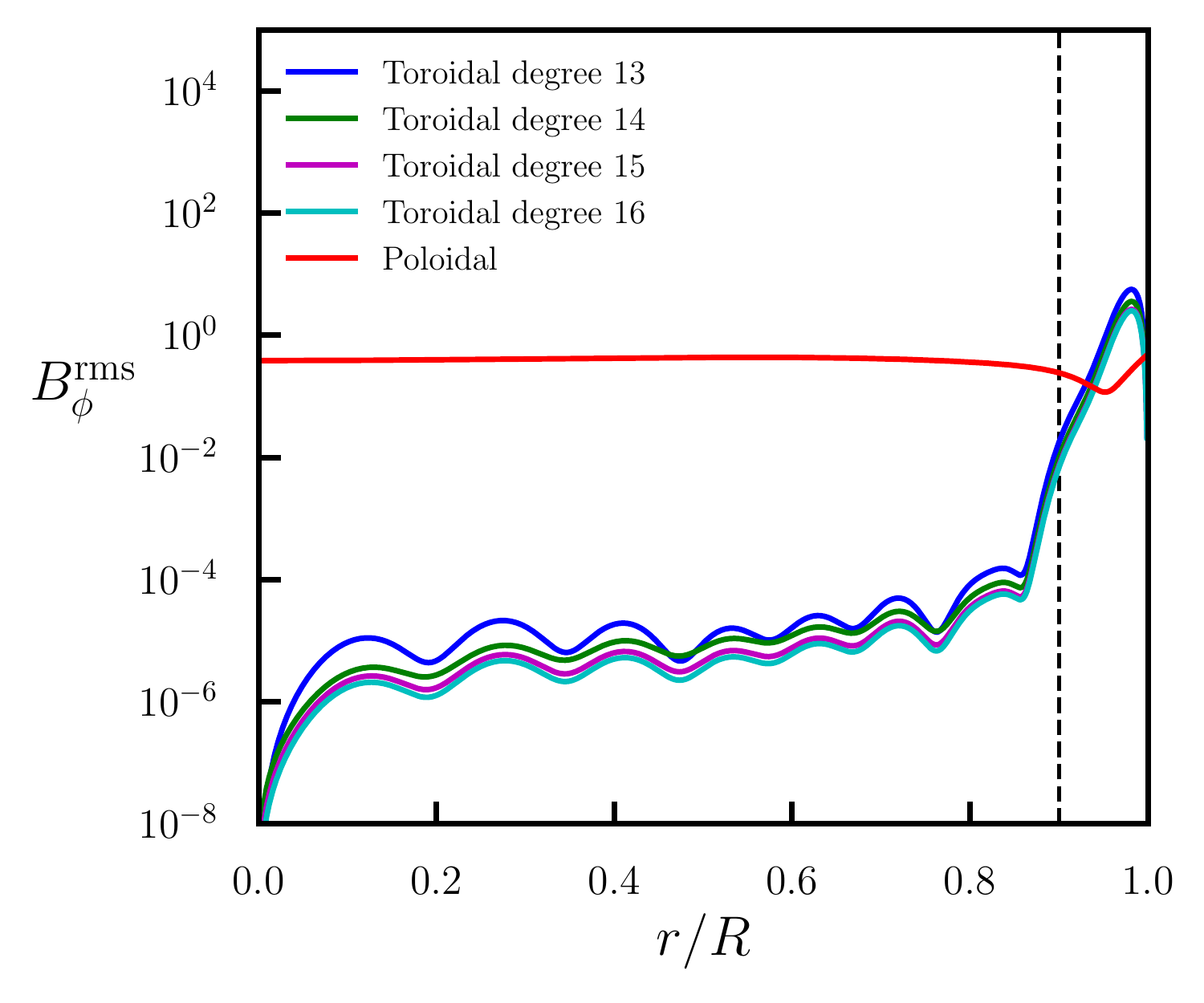}
  \caption{\label{fig:Tor_Pol_field_radial_profile} The root mean squared azimuthal field strength (defined over solid angle) as a function of radius, comparing the strengths of the poloidal field (red) and toroidal field (blue, green, magneta and cyan) for toroidal fields with maximum spherical harmonic degree, order and radial resolution, 13 -- 16 respectively. The poloidal field is the degree 13 field of minimum Ohmic dissipation compatible with the CHAOS-6 model at epoch 2015 \citep{Finlay_etal_2016}. }
  \end{figure}

\Cref{fig:Tor_Pol_field_radial_profile} summarises the strength of toroidal field (in terms of its azimuthal root mean squared value over solid angle) as a function of radius, for different toroidal truncations $L_{max} = N_{max}$ (shown in different colours). 
The toroidal field is required to be four orders of magnitude stronger in the stratified layer in order to satisfy the more restrictive Malkus constraints, compared with the inner region in which the weaker Taylor constraint applies, and adopts a profile that is converged by degree 13.
The strong toroidal field throughout the stratified layer occurs despite the electrically insulating boundary condition at the outer boundary that requires the toroidal field to vanish. 
Within the stratified layer, the azimuthal toroidal field strength attains a maximum rms value of 2.5 mT at a radius of about $0.98R$ or a depth of about 70 km, about double the observed value at the CMB, and locally exceeds the imposed azimuthal poloidal magnetic field (of rms $0.28$ mT at this radius).

\begin{figure}[H]
	\centering
		\begin{subfigure}{.47\textwidth}
		\centering
		\includegraphics[width=.85\linewidth]{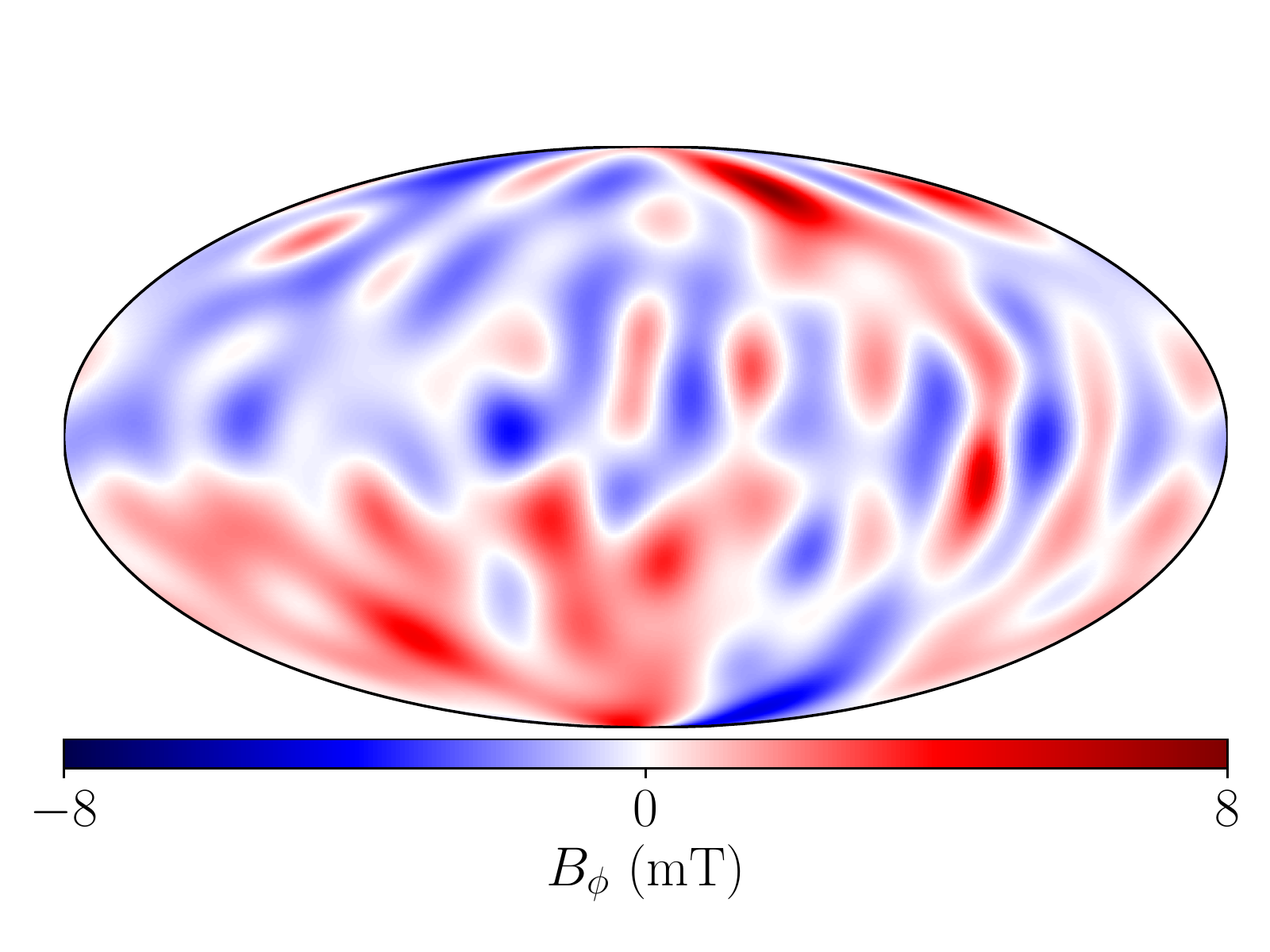}
          \caption{Toroidal field $B_\phi$ at $r=0.98R$, \\(max value = 7.74 mT, RMS = 2.50 mT).}
	\end{subfigure}%
	%
	\begin{subfigure}{.47\textwidth}
	\centering
	\includegraphics[width=.85\linewidth]{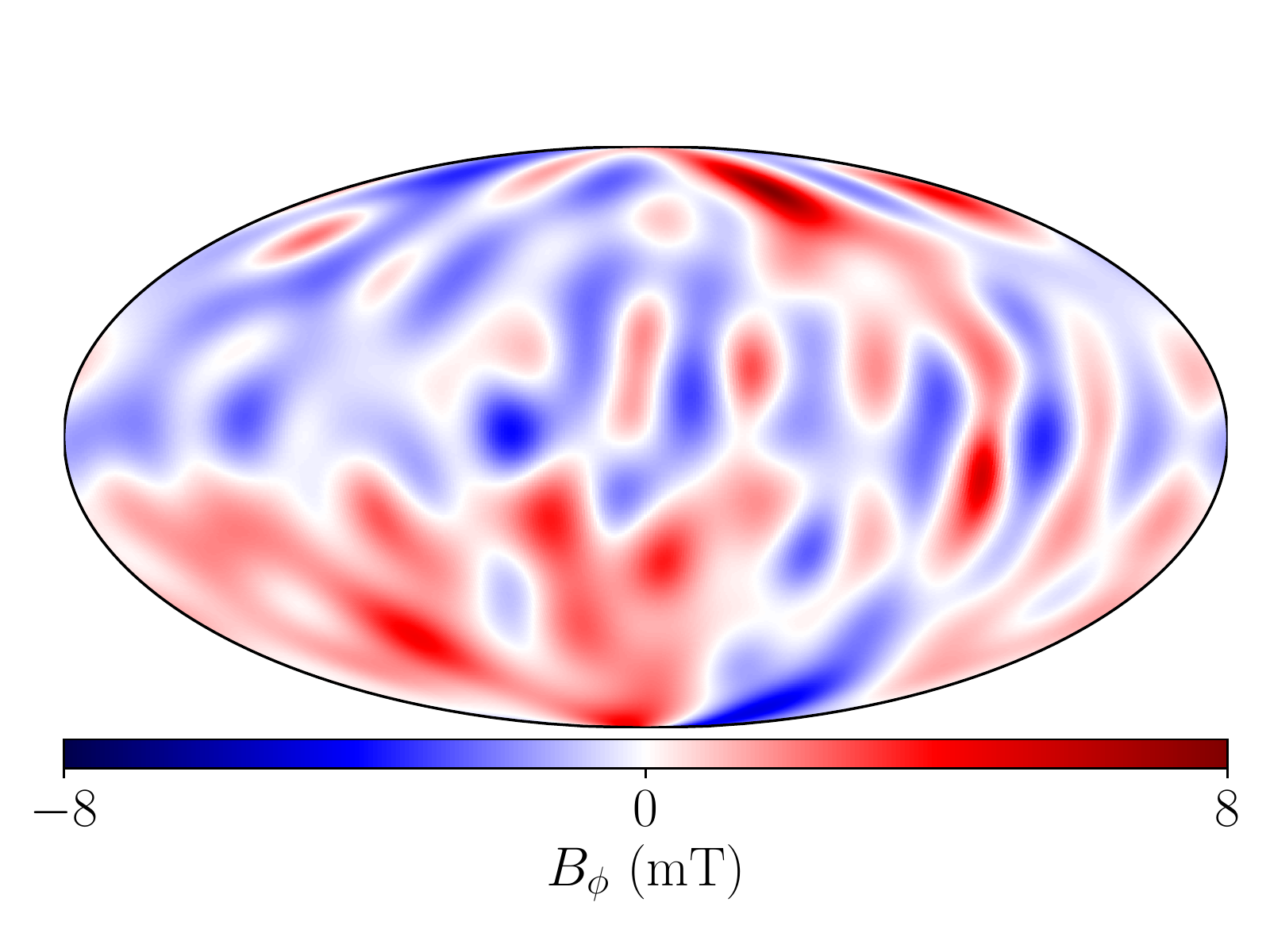}
 \caption{Total field $B_\phi$ at $r=0.98R$, \\(max value = 7.80 mT, RMS = 2.53 mT).}
 \end{subfigure}
 \begin{subfigure}{.47\textwidth}
		\centering
		\includegraphics[width=.85\linewidth]{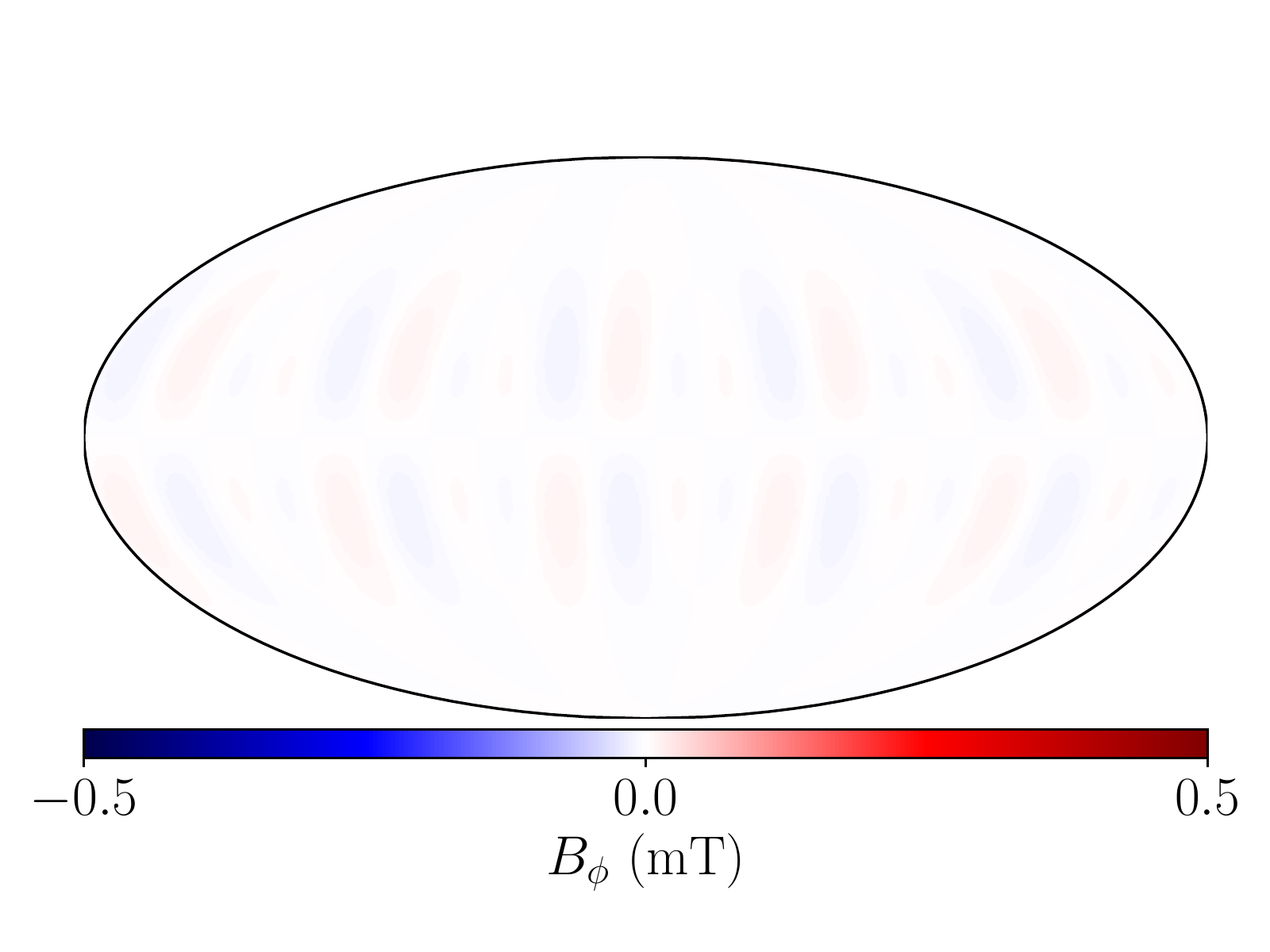}
          \caption{Toroidal field $B_\phi$ at $r=0.7R$, \\(max value = 0.012 mT)}
	\end{subfigure}%
	\begin{subfigure}{.47\textwidth}
	\centering
	\includegraphics[width=.85\linewidth]{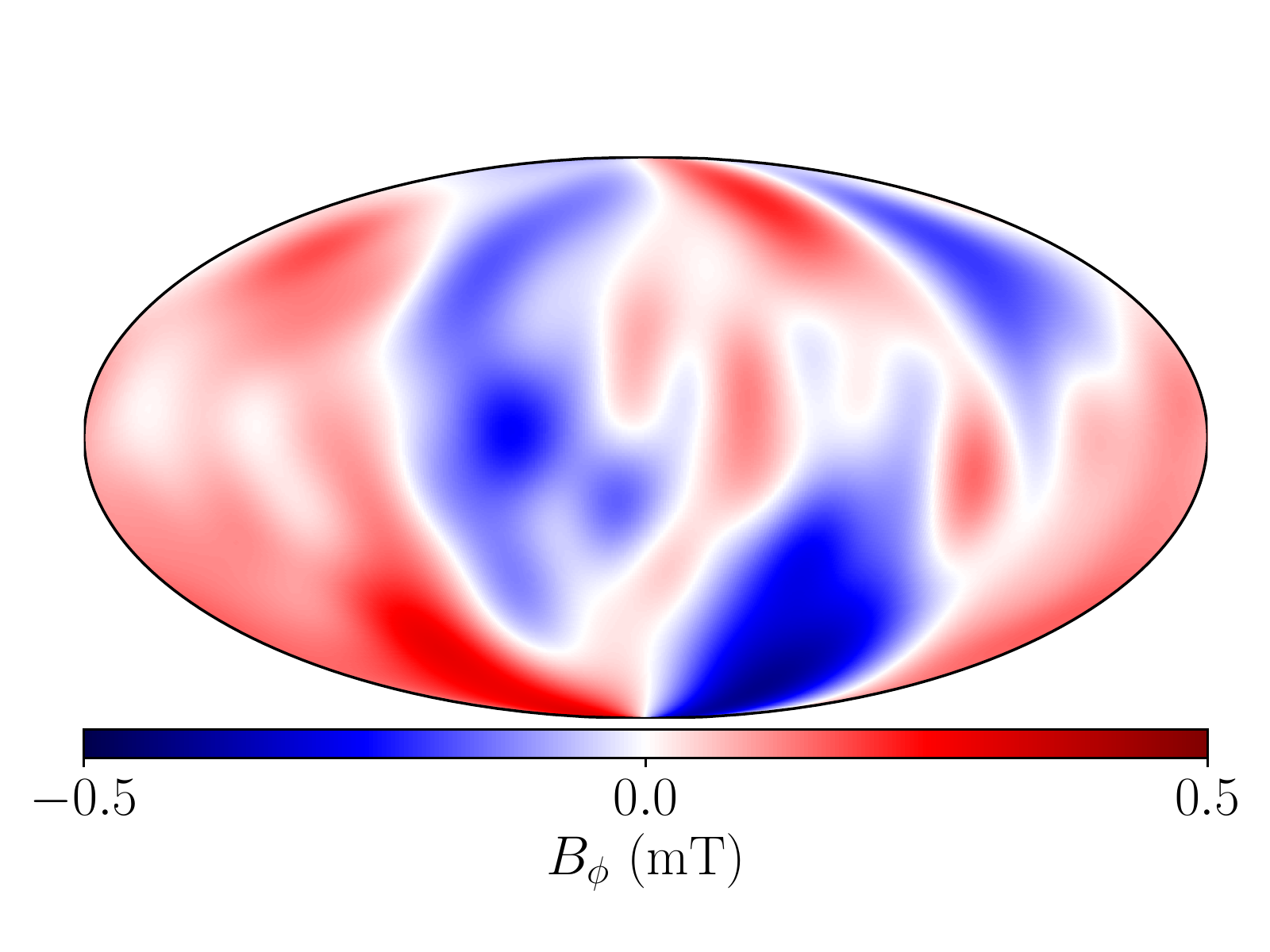}
 \caption{Total field $B_\phi$ at $r=0.7R$, \\(max value = 0.41 mT)}
 \end{subfigure}
 \caption{\label{fig:fullsolution}
 Minimal toroidal-energy solution (a,c) shown by the azimuthal component, of a Malkus state ($0.9R < r \leq R$) and Taylor state $r \le 0.9R$, compared with the total azimuthal component (b,d). Figures (a,b) show the field at a radius of $r=0.98R$, close to where the maximum rms azimuthal toroidal field occurs, while (c,d) show the inner region at $r=0.7R$.}
  \end{figure}

\Cref{fig:fullsolution} shows $B_\phi$ for both the total field and the toroidal component in isolation, using a toroidal truncation of 13 (corresponding to the blue line in \cref{fig:Tor_Pol_field_radial_profile}.)
The top row shows the structure at the radius of maximum rms toroidal field ($r=0.98R$), demonstrating that the additive toroidal field component (of maximum 8 mT) dominates the total azimuthal field.
The bottom row shows a comparable figure at $r=0.7R$, in the inner region where only Taylor's constraint applies. Plotted on the same scale, the required additive toroidal component is tiny compared with the imposed poloidal field. This highlights again that the Malkus constraint is much more restrictive than the Taylor constraint.

\begin{figure}[H] 
	\centering
	\begin{subfigure}{.47\textwidth}
		\centering
		 \includegraphics[width=0.85\textwidth]{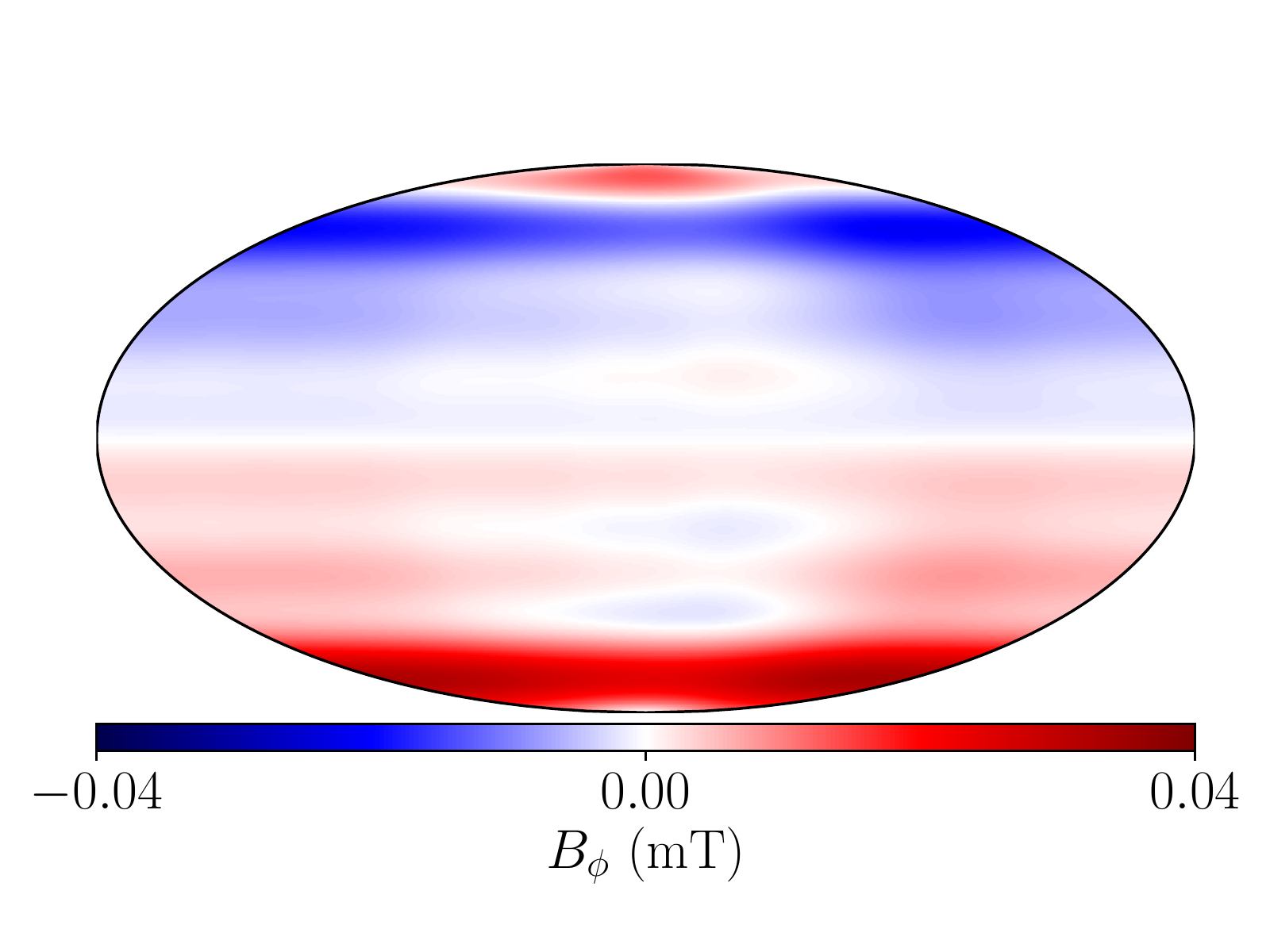}
 \caption{Toroidal field $B_\phi$ at $r=0.98R$, \\(max value = 0.034 mT)\label{fig:Taylor_linear_nonaxi_B_r_r=09_toroidal}}

	\end{subfigure}%
	\begin{subfigure}{.47\textwidth}
	\centering
		\includegraphics[width=.85\linewidth]{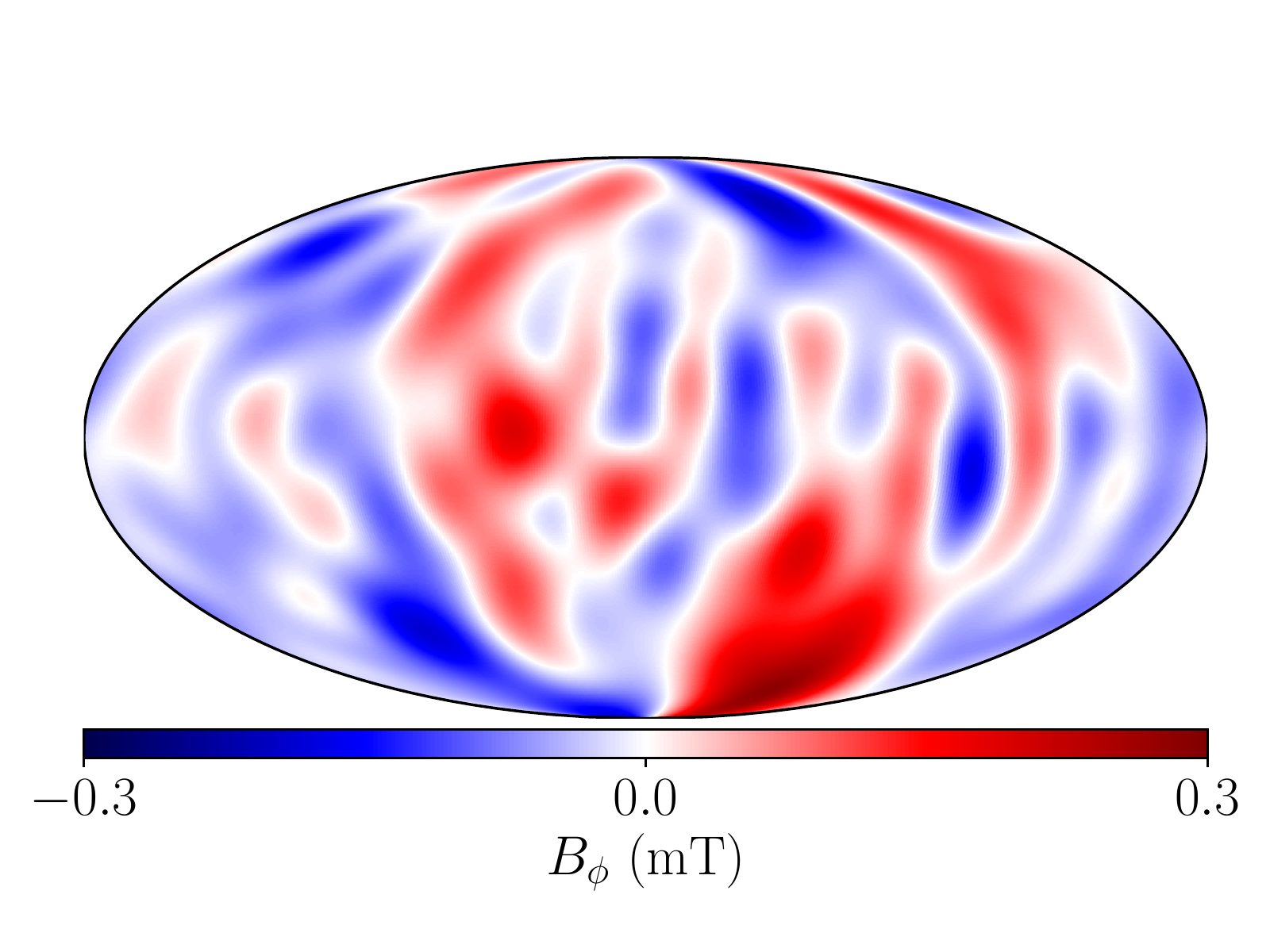}
		  \caption{Total field $B_\phi$ at $r=0.98R$, \\(max value = 0.29 mT) \label{fig:Taylor_linear_nonaxi_B_phi_r=09_poloidal}}
 \end{subfigure}
  \caption{Azimuthal field for an unstratified comparative case, for which the magnetic field satisfies only Taylor's constraint. \label{Tay_fullsphere}}
   \end{figure}

For comparison, \cref{Tay_fullsphere} shows an equivalent solution to \cref{fig:fullsolution}(a,b) but in the absence of stratification (where the magnetic field satisfies only Taylor's constraint).
The toroidal contribution to the azimuthal field is very weak (note the colourbar range is reduced from that of \cref{fig:fullsolution}(a,b) from 8 to 0.04 mT) and is of very large scale.
This further highlights the weakness of the Taylor constraints compared with the Malkus constraints.

   \subsection{Time averaged field over the past ten millenia}
   
Here we show results for a poloidal field that is derived from the 10000-year time averaged field from the CALS10k.2 model \citep{constable2016persistent}. The model is only available up to spherical harmonic degree 10, hence we adopt a truncation of $L_{max} = N_{max} = 10$ for the toroidal field.  Due to the absence of small-scale features in the field (caused by regularisation) the maximum value of $B_r$ is reduced to about $1/2$ of the comparable value from the degree-13 CHAOS-6 model from epoch 2015, and similarly the azimuthal field to about $1/6$ of its value. We note that over a long enough time span, the magnetic field is generally assumed to average to an axial dipole: a field configuration that is both a Malkus state and one in which the azimuthal component vanishes. Thus a small azimuthal component is consistent with such an assumption.
  
  \begin{figure}[H] 
	\centering
	\begin{subfigure}{.47\textwidth}
		\centering
	\includegraphics[width=0.9\textwidth]{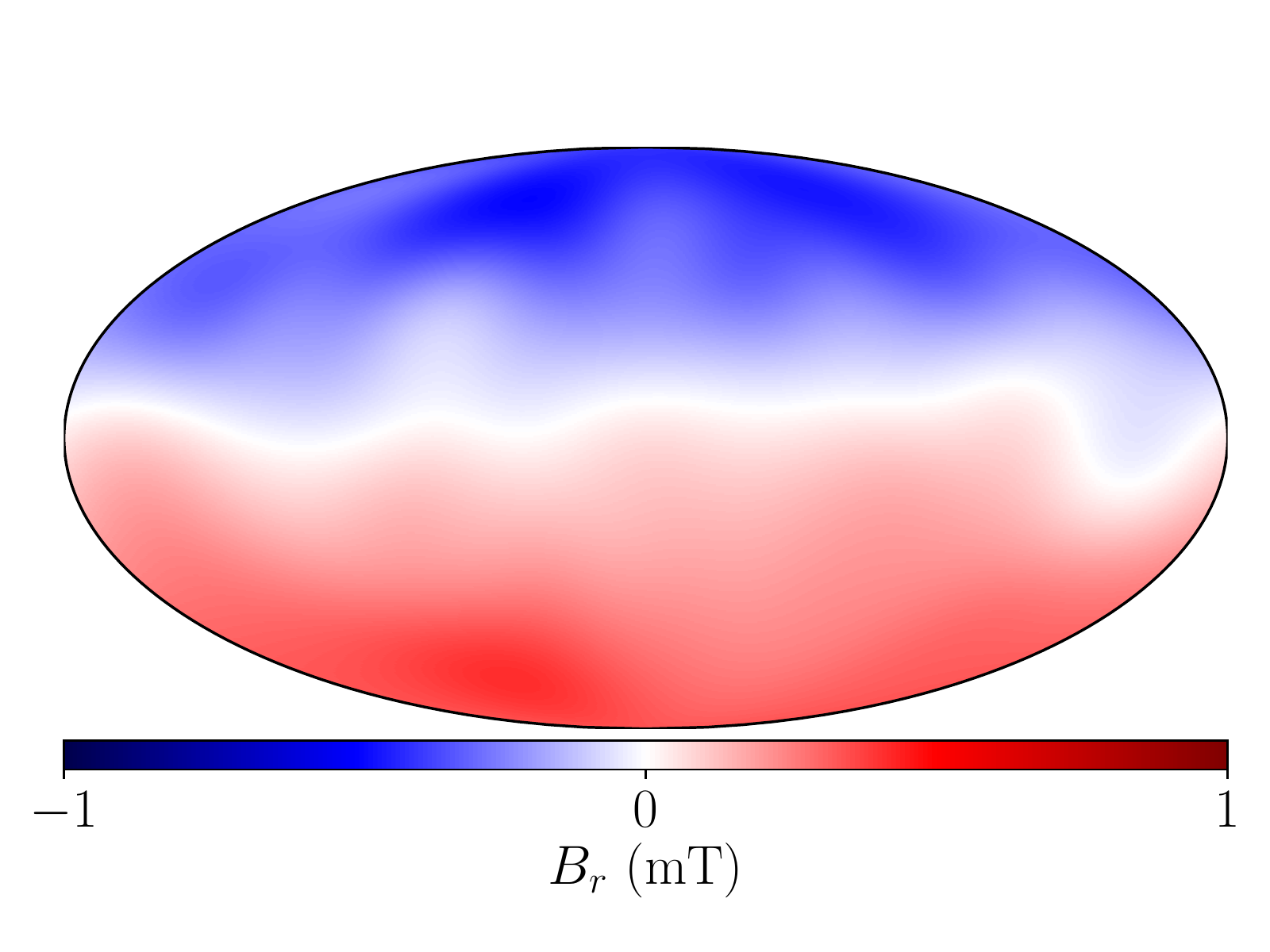}
      \caption{\label{fig:CALSK_10k_CMB_field_basemap} $B_r$, max value = 0.50 mT. }
	\end{subfigure}%
	\begin{subfigure}{.47\textwidth}
	\centering
 \includegraphics[width=0.9\textwidth]{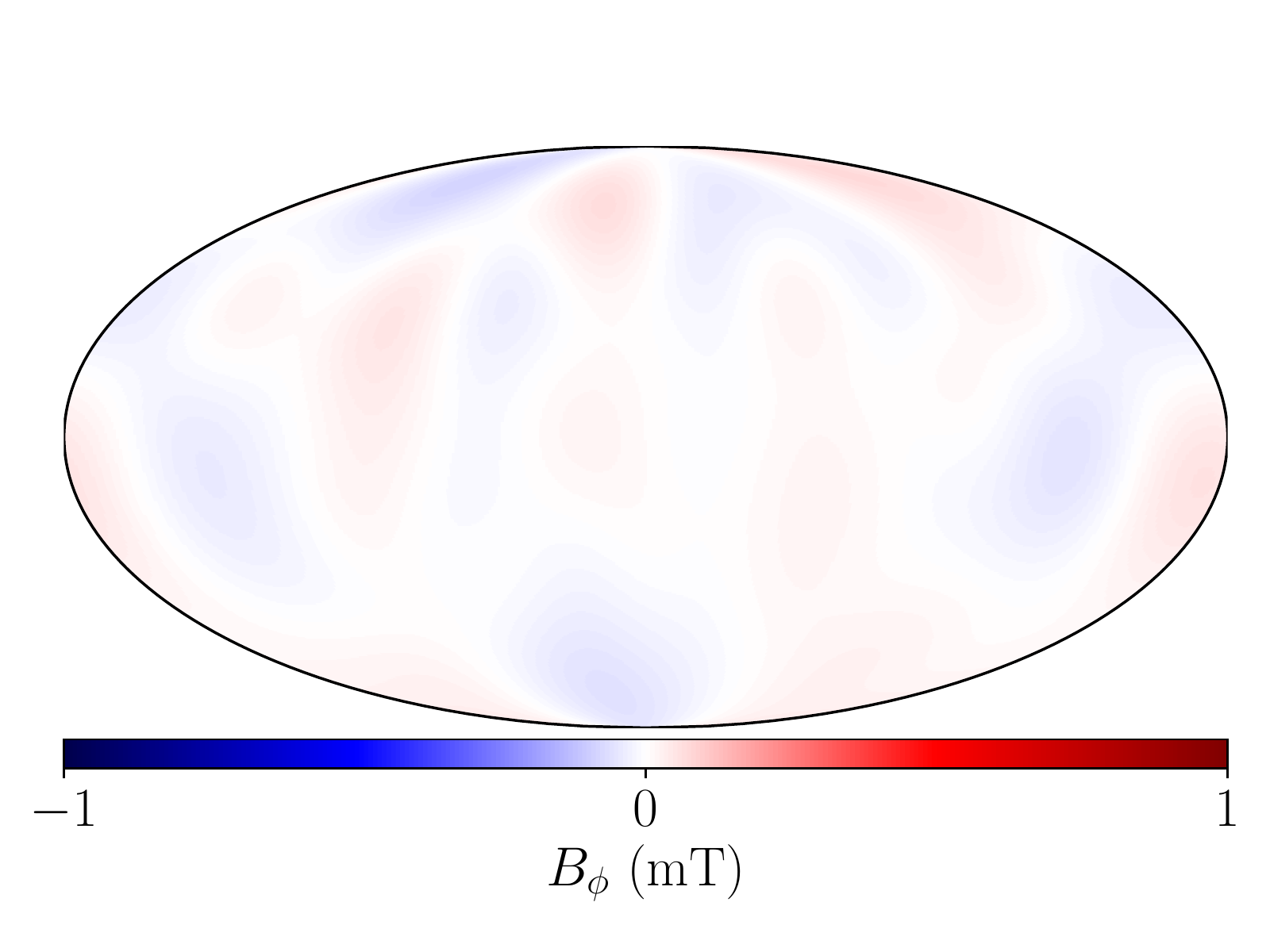}
  \caption{\label{fig:CMB_field_basemap_rot_phi_CALS_phi_new} $B_\phi$, max value = 0.085 mT. }
 \end{subfigure}
 \caption{Magnetic field at the CMB based on the 10000-year time average field from CALS10k.2 \label{fig:CALS_CMB_field_basemap_cor_both}}

   \end{figure}
   
\begin{figure}[H] 
	\centering
	\begin{subfigure}{.47\textwidth}
		\centering
	\includegraphics[width=0.9\textwidth]{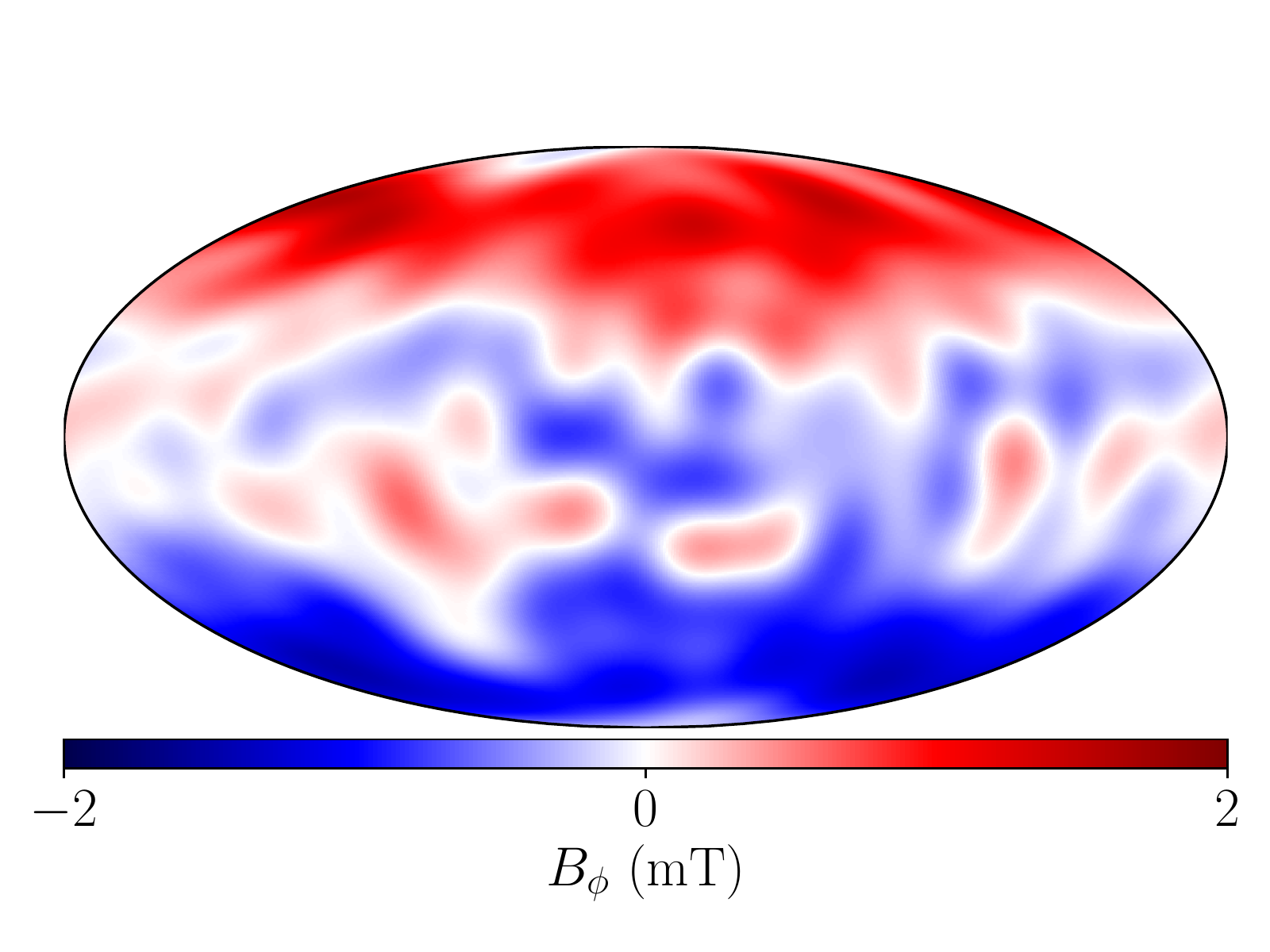}
      \caption{\label{fig:CALS_Min_tot_r095_phi_rot} Toroidal field $B_\phi$ at $r=0.97 R$, \\(max value = 1.60 mT) }
	\end{subfigure}%
	\begin{subfigure}{.47\textwidth}
	\centering
 \includegraphics[width=0.9\textwidth]{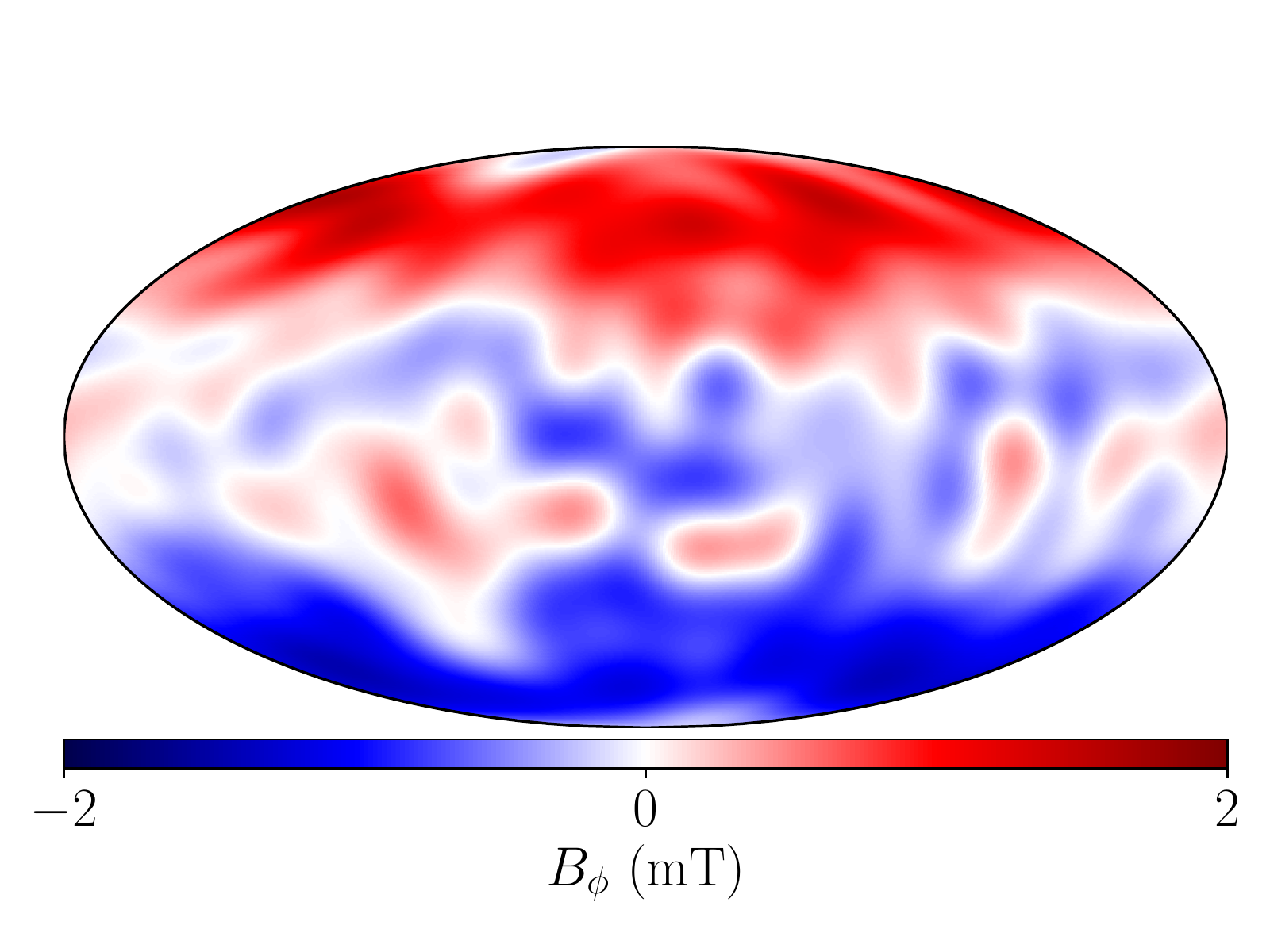}
  \caption{\label{fig:CALS_Min_total_r095_phi_rot} Total field $B_\phi$ at $r=0.97 R$, \\(max value = 1.66 mT) }
 \end{subfigure}
 \caption{The azimuthal component of the Malkus state magnetic field within the stratified layer at a radius of $r=0.97R$, approximately the radius of maximum rms azimuthal toroidal field. 
 \label{fig:CALS_Min_total_r095_phi_rot_both}}
   \end{figure}

   
Contours of the azimuthal field within the stratified layer (at $r=0.97R$) are shown in \cref{fig:CALS_Min_total_r095_phi_rot_both}, which is approximately the radius at which the maximum rms azimuthal toroidal field occurs. As before, the toroidal field dominates the azimuthal component, and its rms (1.66 mT) is about double that on the CMB (0.085 mT).  Although its maximum absolute value is about 3 mT, less than the 8 mT found in the 2015 example above, this is consistent with the overall reduction in structure of the imposed poloidal field.

\section{Discussion}

\subsection{Estimates of the internal magnetic field strength} \label{sec:tor_min}



 %
 Estimating the magnetic field strength inside the core is challenging, because observations made on Earth's surface, using a potential-field extrapolation, only constrain the poloidal magnetic structure down to the CMB and not beneath, but also even this structure is visible only to about spherical harmonic degree 13.  Furthermore, within the framework of such an extrapolation, the toroidal field is zero on the CMB. Estimating the field within the core beyond these surface values requires insight from numerical models or observations of physical mechanisms that are sensitive to the interior field.
 
Based on numerical models,  \cite{Zhang_Fearn_93} suggest that a criterion for stability of the geomagnetic field is a toroidal field no more than 10 times that of the poloidal field, resulting an approximate upper bound of 5 mT. In a more recent and conflicting study, 
  \cite{sreenivasan2017damping} suggest that the mean toroidal field is approximately 10 mT or higher, since this intensity is required for the slow magnetostrophic waves present to be able to originate from small-scale motions in the core.
  Observation-based studies based on electric field measurements \cite{shimizu1998observational} suggest a toroidal field strength at the CMB of anywhere in the range of 1-100 times that of the poloidal magnetic field there (i.e. up to about 100 mT).
  \cite{Buffett_2010} calculated a core averaged field strength of 2.5 mT from measurements of tidal dissipation; \citet{gubbins2007geomagnetic} estimated the toroidal field strength of 1 mT as compatible with patches of reversed magnetic flux. Lastly, the magnetic signature of both torsional and Rossby waves have led to respective estimates of at least 2 mT for $B_s$ within the core and therefore an RMS strength of $4$ mT assuming isotropy \citep{gillet2010fast}, and an RMS estimate of $B_\phi$ of 12 mT \citep{Hori_etal_2015}.

The strong toroidal field within our 2015 models of up to 8 mT (and rms $B_\phi$ of $2.5$ mT) within the stratified layer (at radius $r=0.98R$ or a depth of about 70 km) is in agreement with the majority of these estimates. This maximum value is notably about 8 times stronger than the observed radial field on the CMB.
In both the 2015 and the 10000-yr averaged model, the rms toroidal field within the stratified layer was about double the radial field on the CMB.
Interestingly, the azimuthal component of our solution within the inner unstratified region is about 100 times weaker, demonstrating the extent to which Malkus' constraint is far more restrictive than Taylor's constraint. 

\subsection{Limitations of our model}
Our model does not produce a formal lower bound on the azimuthal component of a magnetic field that (a) satisfies both the Malkus and Taylor constraints in their relevant regions along with (b) constraints on the radial field at the CMB. Instead, our results give only an upper bound on the lower bound \citep[e.g.][]{jackson2011ohmic} because we have made a variety of simplifying assumptions, the most notable of which are (i) we have restricted ourselves to a subspace of Malkus states for which the constraints are linear (ii) we have imposed the entire poloidal profile and (iii) we have used a regular basis set for all magnetic fields even within the stratified layer when this is not strictly necessary. 
However, we show for the example considered in \cref{sec:Ap_sol_simp} that in this case assumption (i) does not have a significant impact and our estimate is close to the full nonlinear lower bound. It may be that the other assumptions also do not cause our azimuthal field estimates to deviate significantly from the actual lower bound. 

Leaving aside the minimum toroidal field suggested by our model, our analysis allows two statements to be made on the weakness of the Malkus constraints, and the ability of magnetic structures assumed on the CMB to probe the magnetic structure within the stratified layer. Firstly, our method can find a toroidal field that converts any poloidal field into a Malkus state within a stratified layer of any depth. This means that we cannot use consistency of observation-derived models of the radial field with the Malkus constraints as a discriminant to test the probe the existence (and depth) of a stratified layer: all such models are consistent. 
%
%
%
Second, even if a stratified layer is assumed, the lateral radial magnetic field structure at the bottom of the layer is unconstrained by its structure at the top because we can find a Malkus state assuming any poloidal profile within the layer. Thus using considerations of the Malkus constraints, models of the surface magnetic field, such as CHAOS-6, cannot be downwards-continued further than the CMB into a stratified layer beneath.

\subsection{Model robustness}

There remains much uncertainty over the depth of any stably stratified layer at the top of the Earth's core \citep{hardy2019stably}. Hence it is natural to consider how our results may change if the layer were to be of a different thickness to the $10\%$ of core radius used, as such we also calculated minimum toroidal-energy solutions matched to CHAOS-6 in epoch 2015 for layer thicknesses of $5\%$ and $15\%$. We find very little dependence of the field strengths internal to the layer on the depth of the layer itself, with our root mean square azimuthal field taking peak values of 2.7, 2.5 and 2.4 mT for thicknesses of $5\%$, $10\%$ and $15\%$ respectively. 

The resolution of poloidal field also impacts significantly our optimal solutions. This has already been identified in the comparison between the degree-13 2015 model, and the degree-10 10,000-yr time-averaged model, that respectively resulted in rms azimuthal field estimates of 2.5 and 1.2 mT.  Interestingly, for very long time-averaging windows the magnetic field is widely supposed to converge to an axial dipole, and assuming a simple poloidal profile is itself an exact Malkus state, with zero azimuthal field strength. 
We can further test 
the effect of resolution by considering
 maximum poloidal degrees of 6 and 10 for the 2015 model to compare with our solution at degree 13. We find that our estimates for the root mean square azimuthal field (taken over their peak spherical surface) were 1.6 and 2.2 mT respectively.  In all these calculations, the spherical harmonic degree representing the toroidal field was taken high enough to ensure convergence. 
Thus stronger toroidal fields are apparently needed to convert more complex poloidal fields into a Malkus state. 
This has important implications for the Earth, for which we only know the 
degree of the poloidal field to about $13$ due to crustal magnetism. Our estimates of the azimuthal field strength would likely increase if a full representation of the poloidal field were known. 

\subsection{Ohmic dissipation}

Our method can be readily amended to minimise the toroidal Ohmic dissipation, rather than the toroidal energy. In so doing, we provide a new estimate of the lower bound of Ohmic dissipation within the core.
%
Such lower bounds are useful geophysically as they are linked to the rate of entropy increase within the core, which has direct implications for: core evolution, the sustainability of the geodynamo, the age of the inner core and the heat flow into the mantle \citep{jackson2011ohmic}.
%

The poloidal field with maximum spherical harmonic degree 13 that we use, based on CHAOS-6 \citep{Finlay_etal_2016} and the minimum Ohmic dissipation radial profile \citep{Book_Backus_etal_96} has by itself an Ohmic dissipation of 0.2 GW. 
\cite{Jackson_Livermore_2009} showed that by adding additional constraints for the magnetic field, a formal lower bound on the dissipation could be raised to 10 GW, and even higher to 100 GW with the addition of further assumptions about the geomagnetic spectrum. This latter bound is close to typical estimates of 1 - 15 TW  \cite{Jackson_Livermore_2009,jackson2011ohmic}. 

The addition of extra conditions derived from the assumed dynamical balance, namely Taylor constraints, were considered by \citet{jackson2011ohmic} by adopting a very specific magnetic field representation. These constraints alone raised their estimate of the lower bound from 0.2 to 10 GW, that is, by a factor of 20.
In view of the much stronger Malkus constraints (compared to the Taylor constraints), we briefly investigate their impact here.

We follow our methodology and find an additive toroidal field of minimal dissipation (rather than energy) that renders the total field a Malkus state. The dissipation is altered from $0.2$ to 
$0.7$ GW. That this increase is rather small (only a small factor of about 3) is rather disappointing, but is not in contradiction to our other results. It is generally true that the Malkus constraints are more restrictive than the Taylor constraint, but this comparison can only be made when the same representation is used for both. The method of \citet{jackson2011ohmic} assumed a highly restrictive form, so that in fact their Taylor states were apparently actually more tightly constrained than our Malkus states and thus produced a higher estimate of the lower bound.  
Despite our low estimate here, additional considerations of Malkus constraint may increase the highest estimates of \citet{Jackson_Livermore_2009} well into the geophysically interesting regime.


%


\subsection{Further extensions}

The Malkus states we have computed, which match to field observations, provide a plausible background state at the top of the core. 
It may be interesting for future work to investigate how waves thought to exist within such a stratified layer \citep{Buffett_2014} may behave when considered as perturbations from such a background state, and whether they remain valid suggestions for explaining secular variation in the geomagnetic field. Similarly, combining our analysis with constraints on $B_s$ from torsional wave models \citep{gillet2010fast} may be insightful, and would combine aspects of both long and short-term dynamics.
%
%

It is worth noting though, that we have investigated only static Malkus states without consideration of dynamics: we do not require the magnetic field to be either steady or stable, both of which would apply additional important conditions. An obvious extension to this work then is to investigate the fluid flows which are generated by the Lorentz force associated with these fields. This would then allow a consideration of how such flows would modify the field through the induction equation.
These dynamics are however, are still relatively poorly known even for the much simpler problem of Taylor states. Recent progress by \cite{hardy2018determination} now allows a full calculation of the flow driven by a Taylor state. A general way to discover stationary and stable Taylor states comparable with geomagnetic observations is still out of reach, and currently the only way to find a stable Taylor state is by  time-stepping \citep[e.g.][]{li2018taylor}.

The well established test used to determine whether the appropriate magnetostrophic force balance is achieved within numerical dynamo simulations is `Taylorisation', which represents a normalised measure of the magnitude of the Taylor integral \cref{eqn:Taylor} and hence the departure from the geophysically relevant, magnetostrophic regime \citep[e.g.][]{Takahashi_etal_2005}. 

We propose an analogous quantity termed `Malkusisation' defined in the same way, in terms of the Malkus integral:

$$\text{Malkusization} = \frac{|\int_0^{2\pi} ([\curl \bB] \times \bB)_\phi d\phi|}{ \int_0^{2\pi} | ([\curl \bB] \times \bB)_\phi | d\phi } $$

This quantity is expected to be very small within a stratified layer adjacent to a magnetostrophic dynamo, provided stratification is sufficiently strong. The recently developed dynamo simulations of \cite{olson2018outer,stanley2008effects} which incorporate the presence of a stratified layer can utilise the computation of this quantity to access the simulation regime.

Finally, we note that the appropriate description of a stratified layer may in fact need to be more complex than a single uniform layer that we assume. Numerical simulations of core flow with heterogeneous CMB heat flux by \cite{mound2019regional} find that localised subadiabatic regions that are stratified are possible amid the remaining actively convecting liquid. 
If indeed local rather than global stratification is the more appropriate model for the Earth's outermost core then the condition of requiring an exact Malkus state would not apply, and the constraints on the magnetic field would be weakened by the existence of regions of non-zero radial flow.
\section{Conclusion}

In this paper we have shown how to construct magnetic fields that are consistent with geomagnetic observations, a strongly stratified layer and the exact magnetostrophic balance thought to exist within Earth's core. 

To do this, we derived the Malkus constraints that must be satisfied by such a magnetic field, whose 
structure gives insight into the nature of the Earth's magnetic field immediately beneath the CMB, where a layer of stratified fluid may be present. 
%

For a fixed magnetic field resolution, although the Malkus constraints are more numerous than the Taylor constraints, many solutions compatible with geomagnetic observations still exist. By making further assumptions about the field structure, we estimate that the toroidal field within the stratified layer is about 8 mT, significantly stronger than the 1 mT of the radial field inferred from degree-13 observations. 

\section{Acknowledgements}

This work was supported by the Engineering and Physical Sciences Research Council (EPSRC) Centre for Doctoral Training in Fluid Dynamics at the University of Leeds under Grant No. EP/L01615X/1. P.W.L. acknowledges partial support from NERC grant NE/G014043/1. The authors would also like to thank Dominique Jault and Emmanuel Dormy for helpful discussions, as well as the Leeds Deep Earth group for useful comments. 
Figures were produced using matplotlib \citep{Hunter_2007}.

\appendix

\section{Full sphere Malkus states} \label{sec:Apa_both}

\subsection{Simple Example} \label{sec:Apb}

Here we consider a simple example of an axisymmetric magnetic field in a full sphere of radius $R$, consisting of four modes: a toroidal $l=1$, $n=1$ mode, a toroidal $l=1$, $n=2$ mode, a poloidal $l=1$, $n=1$ mode and a poloidal $l=1$, $n=2$ mode, each of which has an unspecified corresponding coefficient $\alpha_{l,n}$ and $\beta_{l,n}$ for toroidal and poloidal modes respectively. 
Through this we demonstrate the form of the linear constraints which arise from Malkus' constraint in this case. It is significant to note the vital role of degeneracy within these constraints in permitting a solution.

Through computing the Malkus integral and enforcing that this is zero for all values of $s$ and $z$ by requiring that the coefficients of all powers of $s$ and $z$ vanish we obtain a series of simultaneous equations:

$$\left(-\frac{11}{8}\beta_{1,2}+2\beta_{1,1}\right)\alpha_{1,2}-\frac{253}{140}\left(\frac{77}{69}\beta_{1,2}+\frac{56}{759}\beta_{1,1}\right)\alpha_{1,1}=0,$$

$$\left(\frac{319}{84}\beta_{1,2}-\frac{10}{3}\beta_{1,1}\right)\alpha_{1,2}-\frac{253}{70}\beta_{1,2}\alpha_{1,1}=0,$$

$$\left(-\frac{165}{56}\beta_{1,2}+\beta_{1,1}\right)\alpha_{1,2}-\frac{253}{140}\beta_{1,2}\alpha_{1,1}=0,$$

$$\left(\frac{319}{84}\beta_{1,2}-\frac{10}{3}\beta_{1,1}\right)\alpha_{1,2}-\frac{253}{70}\beta_{1,2}\alpha_{1,1}=0,$$

$$\left(-\frac{165}{28}\beta_{1,2}+2\beta_{1,1}\right)\alpha_{1,2}-\frac{253}{70}\beta_{1,2}\alpha_{1,1}=0,$$

$$\left(-\frac{165}{56}\beta_{1,2}+\beta_{1,1}\right)\alpha_{1,2}-\frac{253}{140}\beta_{1,2}\alpha_{1,1}=0.$$

Although there are 6 equations here, there are only two independent conditions:

$$\alpha_{1,1}\beta_{1,2}+\frac{5}{2}\alpha_{1,2}\beta_{1,2}=0, ~~~~ \text{and} ~~~~ \alpha_{1,2}\beta_{1,1}+\frac{11}{7}\alpha_{1,2}\beta_{1,2}=0.$$
If both $\beta_{1,2}$ and $\alpha_{1,2}$ are nonzero, then these become linear constraints.

Hence, in this case we can see that there are 4 degrees of freedom, 6 constraint equations but only 2 independent constraints.
This means that while on first inspection the system appears to be overconstrained with no solution, there are in fact multiple Malkus state solutions, with the solution space being spanned by two degrees of freedom ($\beta_{1,2}, \alpha_{1,2}$) with the 
other coefficients determined in terms of these by the relationships:

$$\alpha_{1,1}=-\frac{5}{2}\alpha_{1,2} ~~~~ \text{and} ~~~~ \beta_{1,1}=-\frac{11}{7}\beta_{1,2}.$$

Despite the significant degenercy in the Malkus constraints, they are 
notably more restrictive than the Taylor constraints for this truncation of $L_{max}=1$, $N_{max}=2$, for which the Taylor integral is identically zero and so provides no restriction. 

\subsection{Solution of a low-resolution system} \label{sec:Ap_sol_simp}

\begin{figure}[H] 
	\centering
	\begin{subfigure}{.4\textwidth}
		\centering
\includegraphics[width=1\textwidth]{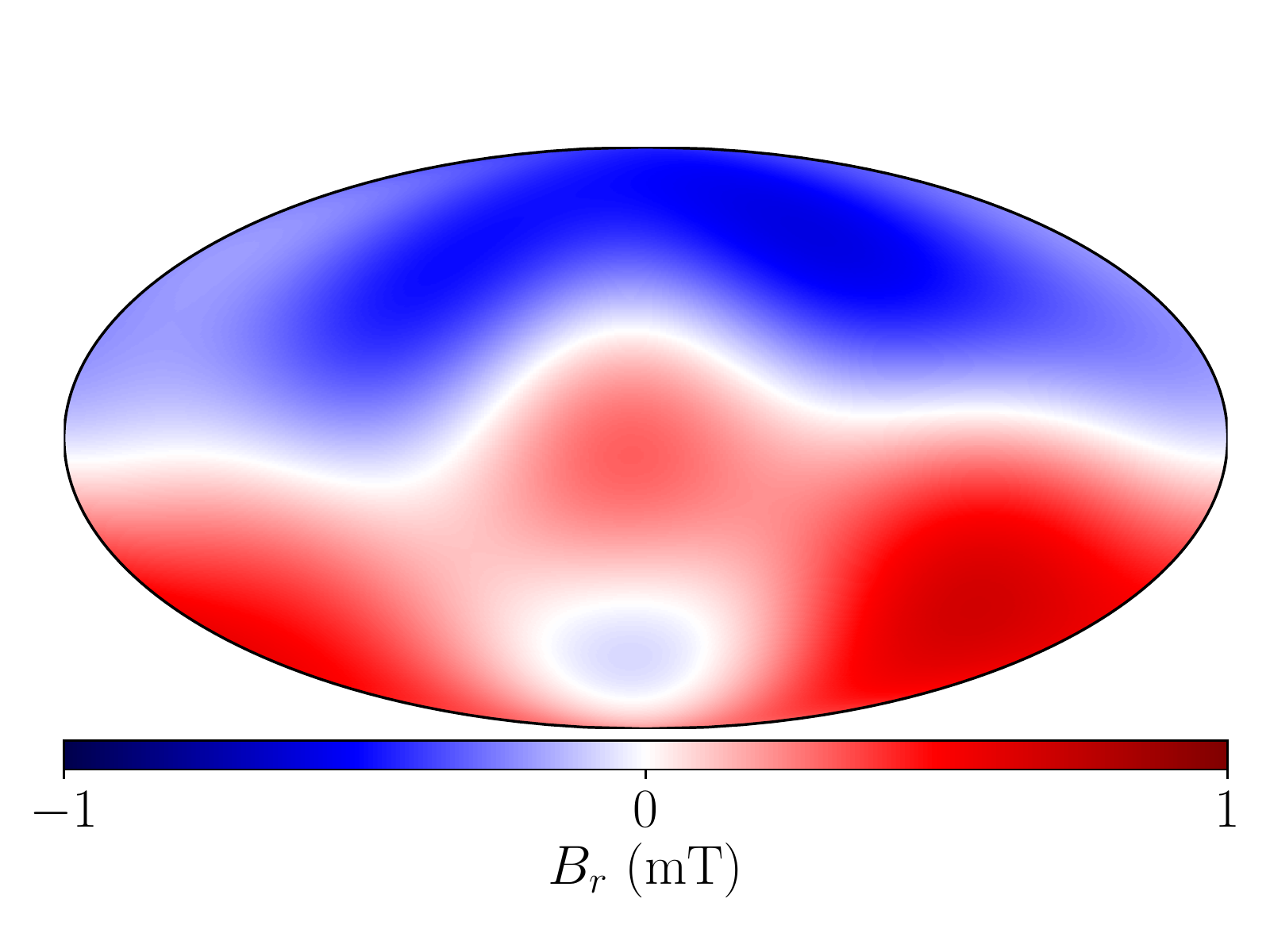}
\caption{ $B_r$ at $r=0.9R$ \label{fig:r09_fixpol}}
	\end{subfigure}%
	\begin{subfigure}{0.4\textwidth}
	\includegraphics[width=1\textwidth]{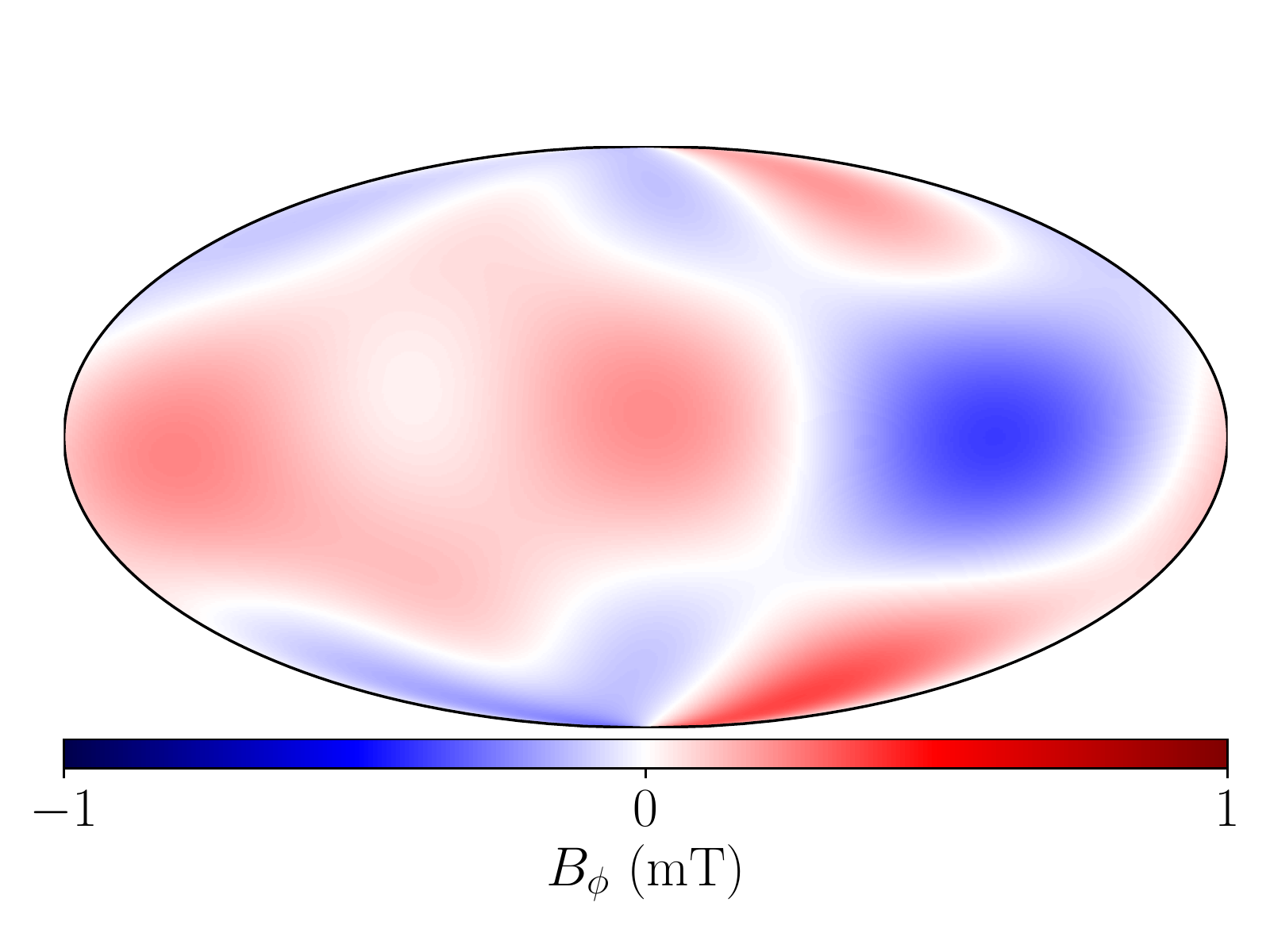}
 \caption{$B_\phi$ at $r=0.9R$ \label{fig:phi_r09_fixpol}}
	\end{subfigure}
	\begin{subfigure}{.4\textwidth}
		\centering
\includegraphics[width=1\textwidth]{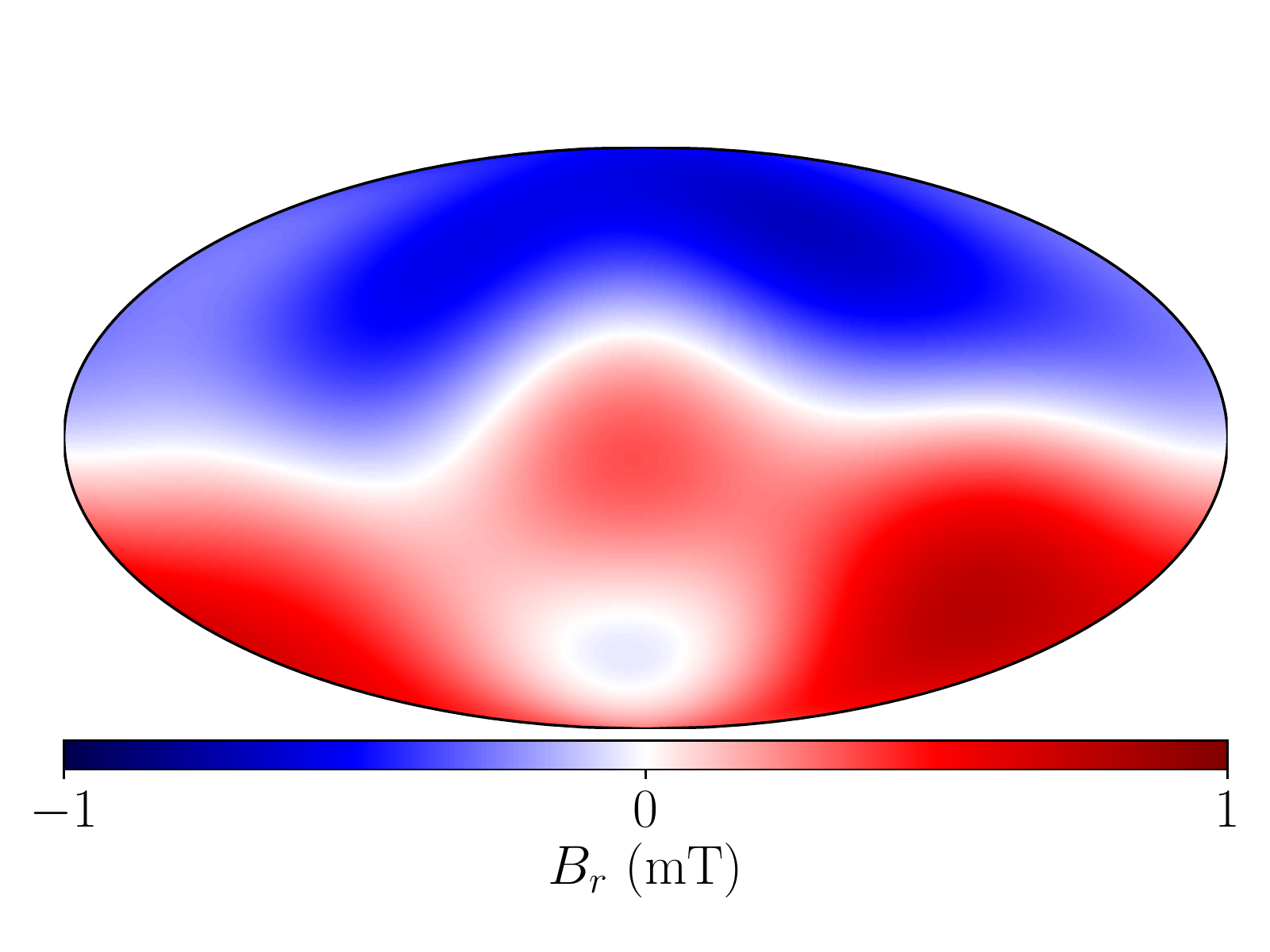}
\caption{ $B_r$ at $r=0.8R$ \label{fig:r08_fixpol}}
	\end{subfigure}%
	\begin{subfigure}{0.4\textwidth}
	 \includegraphics[width=1\textwidth]{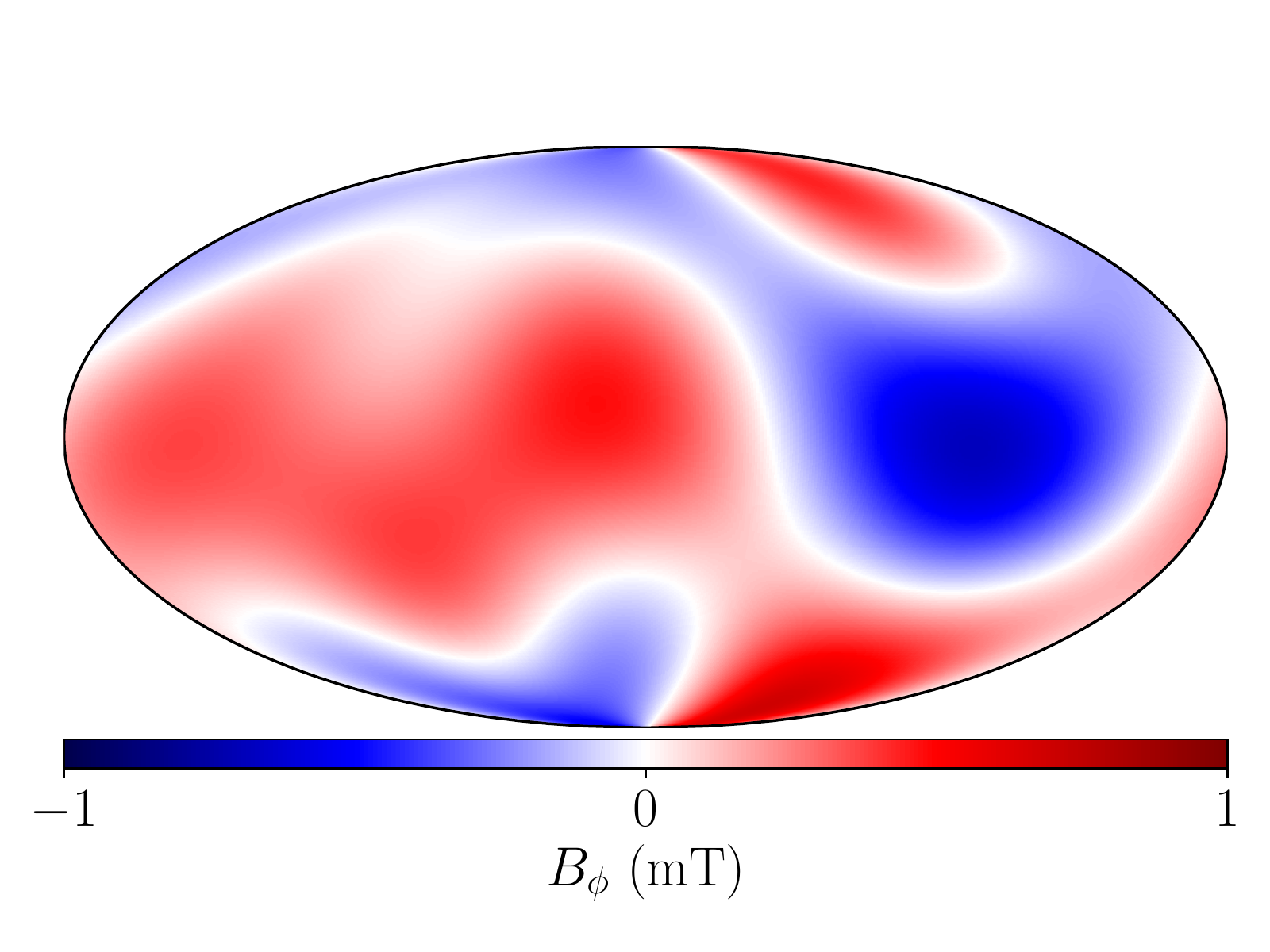}
 \caption{$B_\phi$ at $r=0.8R$\label{fig:phi_r08_fixpol}}
	\end{subfigure}
	
    \caption{\label{fig:fixpol_nonlin_all} Malkus state with $L_{max}=3, ~ N_{max}=3, ~ M_{max}=3$. }
\end{figure}

We now present the first known non-trivial solution of a Malkus state. Here we consider a full sphere magnetic field truncated at $L_{max}=3, ~ N_{max}=3, ~ M_{max}=3$, and impose a minimum Ohmic dissipation poloidal profile that matches the CHAOS-6 model (to degree 3) at $r=R$.
We seek a toroidal field using all spherical harmonic modes within the truncation $L_{max}=3,~N_{max}=3$ (described by 45 degrees of freedom) that when added to this poloidal field satisfies the Malkus constraints.
Of the 72 nonlinear constraint equations, only 42 are independent. Thus the number of degrees of freedom exceed the number of independent constraints, although since the constraints are nonlinear it is not immediate that a solution exists.
However, using the computer algebra software Maple, we find the solution that minimises toroidal field strength as well as satisfying all the constraints, which is visualised in \cref{fig:fixpol_nonlin_all}.
We cannot generalise this procedure to higher resolutions because of the numerical difficulty in finding optimal solutions in such a nonlinear problem.

For comparison we also compute the solution using the method described in \cref{sec:theo}, which owing to the specific choice of toroidal spherical harmonic modes results in a linear system. The qualitative similarity between these solutions is important in giving an insight into how important it is that we make the (necessary) choice for our higher resolution Earth-like solutions, of only including modes that result in a linear system. Quantitatively this holds too, with rms values of $B_\phi$ of 0.21 and 0.23 mT for the non-linear and linear solutions respectively.
Hence we suggest that the estimates for the lower bound of Earth's toroidal field strength we have calculated would not be significantly different were it possible to solve the full non-linear system.

    \begin{figure}[H]
  \centering
  \includegraphics[width=0.4\textwidth]{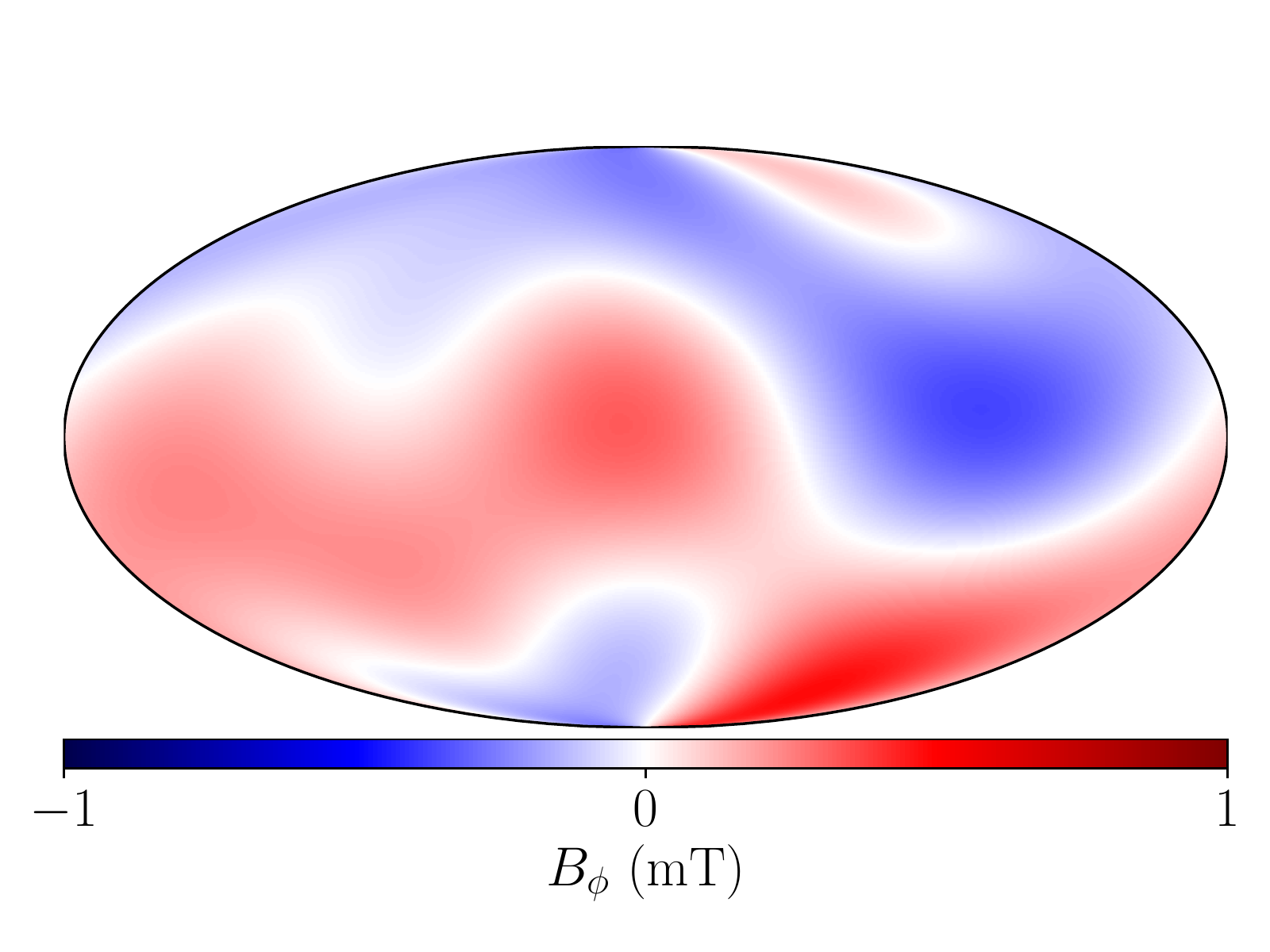}
  \caption{Linear solution for $B_\phi$ at $r=0.9 R$, using the method outlined in \cref{sec:theo} and used for the Earth-like solutions}
  \end{figure}

\section{Enumeration of constraints} \label{sec:enum_con}

In order to determine the number of Malkus constraints, we calculate the maximum possible exponent in dimension of length within the Malkus integral. Since each constraint equation arises from ensuring a coefficient of a different exponent vanishes, enumerating all possibilities gives the maximum number of constraints.


There are three possible non-zero interactions whose sum comprise the Malkus integral, toroidal-toroidal, toroidal-poloidal and poloidal-poloidal as defined in \cref{eq:tortor}.
Since the poloidal field definition contains two curls whereas the toroidal field only one, then this extra derivative reduces the maximum exponent by one for interations involving a poloidal field as opposed to a toroidal one. This means that the maximal case is determined by the toroidal-toroidal interaction, $[\vec{\mathcal{T}_1},\vec{\mathcal{T}}_2]$.
Since the Malkus integrand is identical to the Taylor integrand, we observe that the maximum radial exponent in the Malkus integrand $((\curl \vec{\mathcal{T}_1})\times \vec{\mathcal{T}_2})_\phi$ is  $2L_{max}+4N_{max}-1$, as derived by \cite{livermore2008structure}.
%
%
This is then reduced by two due to the property that the interaction of two toroidal harmonics that have identical spherical harmonic degrees and orders is zero \citep{livermore2008structure}. This requires that one of the two modes has an $L_{max}$ of at least one smaller than the other, hence resulting in a maximum possible degree in $r$ of $2L_{max}+4N_{max}-3$.

Now under a transform in coordinate systems we note that $r^n$ in spherical coordinates can be expressed as $s^jz^k$ in cylindrical coordinates, where $n=j+k$.
Since only even values of $j$ are present this results in $L_{max}+2N_{max}-2 = C_T$ non-trivial constraint equations in this dimension.
%
%
%
%
%
There is no such restriction on $k$, which can take all values up to the maximum of $2L_{max}+4N_{max}-3 = 2C_T+1$.


%
%
%
%
Each one of the constraints arises from a coefficient of a term with a different combination of exponents in $s$ and $z$, explicitly, these terms have the following form: 
\begin{align} &(A_{C_T,0}z^0 + A_{C_T,1}z) s^{2C_T} +(A_{C_T-1,0}z^0+A_{C_T-1,1}z+A_{C_T-1,2}z^2+A_{C_T-1,3}z^3)s^{2(C_T-1)}\nonumber \\&+(A_{C_T-2,0}z^0+\dots+A_{C_T-2,5}z^5)s^{2(C_T-2)} +\dots \nonumber\\& +(A_{1,0}+\dots+A_{1,2C_T-1}z^{2C_T-1})s^2+(A_{0,0}+\dots+A_{0,2C_T+1}z^{2C_T+1}).\end{align}
%
%
%
%
%
%
Hence from the summation of the total number of these terms for every combination of $j$ and $k$, with $j$ even, such that $j+k \leq 2C_T+1$ we have the following expression for the maximum number of Malkus constraints,
$$C_M = 2 \sum_{n=0}^{C_T} (n+1) = (C_T+1)(C_T+2) = C_T^2 + 3C_T + 2. $$




\bibliography{allrefs.bib}

\bibliographystyle{plainnat}

\end{document}